\documentclass[twocolumn,twocolappendix]{aastex631}

\usepackage{lipsum}
\usepackage{amsmath}
\usepackage{enumitem}
\usepackage{booktabs}

\newif\ifGarcia
\Garciatrue

\newif\ifRobinson
\Robinsontrue

\newif\ifedits
\editsfalse
\ifedits
    \renewcommand{\edit}[1]{{\color{red} #1}}
    
\else
    \renewcommand{\edit}[1]{#1}
    
\fi

\begin{document}

\title{Star Formation Rates, Metallicities, and Stellar Masses on kpc-scales in TNG50}

\correspondingauthor{Alex M. Garcia} \\
\email{alexgarcia@virginia.edu}

\author[0000-0002-8831-4922]{Jia Qi}
\altaffiliation{These authors contributed equally to this work}
\affiliation{Department of Astronomy, University of Florida, 
211 Bryant Space Sciences Center,
Gainesville, FL 32611, USA  }

\author[0000-0002-8111-9884]{Alex M. Garcia}
\altaffiliation{These authors contributed equally to this work}
\affiliation{Department of Astronomy, University of Virginia,
530 McCormick Road, 
Charlottesville, VA 22904}

\author[0009-0005-8832-6033]{Davis Robinson}
\affiliation{Department of Astronomy, University of Virginia,
530 McCormick Road, 
Charlottesville, VA 22904}

\author[0000-0002-5653-0786]{Paul Torrey}
\affiliation{Department of Astronomy, University of Virginia, 
530 McCormick Road, 
Charlottesville, VA 22904}

\author[0000-0002-3430-3232]{Jorge Moreno}
\affiliation{Department of Physics and Astronomy, Pomona College, 
Claremont, CA 91711, USA}
\affiliation{Carnegie Observatories, Pasadena, CA 91101, USA}

\author[0009-0002-2049-9470]{Kara N. Green}
\affiliation{Department of Astronomy, University of Virginia, 
530 McCormick Road, 
Charlottesville, VA 22904}

\author[0000-0003-2638-1334]{Aaron S. Evans}
\affiliation{Department of Astronomy, University of Virginia, 
530 McCormick Road, 
Charlottesville, VA 22904}
\affiliation{National Radio Astronomy Observatory, 520 Edgemont
Road, Charlottesville, VA 22903}

\author[0000-0002-0799-0225]{Z. S. Hemler}
\affiliation{Department of Astrophysical Sciences, Princeton University, 
Peyton Hall, 
Princeton, NJ, 08544, USA }

\author{Lars Hernquist}
\affiliation{Center for Astrophysics, Harvard \& Smithsonian, 
60 Garden Street, 
Cambridge, MA 02138, USA}

\author[0000-0002-1768-1899]{Sara L. Ellison}
\affiliation{Department of Physics \& Astronomy, University of Victoria,
Finnerty Road,
Victoria, British Columbia, V8P 1A1, Canada}

\begin{abstract}
Integral field units (IFU) have extended our knowledge of galactic properties to kpc (or, sometimes, even smaller) patches of galaxies.
These scales are where the physics driving galaxy evolution (feedback, chemical enrichment, etc.) take place.
Quantifying the spatially-resolved properties of galaxies, both observationally and theoretically, is therefore critical to our understanding of galaxy evolution.
To this end, we investigate spatially-resolved scaling relations within galaxies of $M_\star>10^{9.0}$ at $z=0$ in IllustrisTNG.
We examine both the resolved star-forming main sequence (rSFMS) and the resolved mass-metallicity relation (rMZR) using $1~{\rm kpc}\times1~{\rm kpc}$ maps. 
We find that the rSFMS in IllustrisTNG is well-described by a power-law, but \edit{is significantly shallower than the observed rSFMS}.
\edit{However, the disagreement between the rSFMS of IllustrisTNG and observations is likely driven by an overestimation of AGN feedback in IllustrisTNG for the higher mass hosts.}
Conversely, the rMZR for IllustrisTNG \edit{has very good agreement with observations.}
Furthermore, we argue that the rSFMS is an indirect result of the Schmidt-Kennicutt (SK) law and local gas relation, which are both independent of host galaxy properties. 
Finally, we expand upon a localized leaky-box model to study the evolution of idealized spaxels and find that it provides a good description of these resolved relations.
The degree of agreement, however, between idealized spaxels and simulated spaxels depends on the `net' outflow rate for the spaxel, and the \edit{IllustrisTNG} scaling relations indicate a preference for a low net outflow rate. 
\end{abstract}


\section{Introduction}
\label{sec:intro}
Scaling relations have long held an important role in describing the nature of galaxies. 
Perhaps unsurprisingly, galaxies do not have randomly assigned properties, but instead follow well-defined relations.
These scaling relations typically compare one integrated property of a galaxy against another (e.g., velocity dispersion and absolute magnitude; \citeauthor{Faber_Jackson_1976} \citeyear{Faber_Jackson_1976}).
However, with the advent of integral field spectroscopy (IFS) surveys (e.g., CALIFA; \citeauthor{Sanchez2012} \citeyear{Sanchez2012}, SAMI; \citeauthor{Bryant2015} \citeyear{Bryant2015}, MaNGA; \citeauthor{Bundy2015} \citeyear{Bundy2015}, PHANGS; \citeauthor{Leroy_2021} \citeyear{Leroy_2021}), these ``global'' scaling relations have been extended down to spatially-resolved scales of $\lesssim1$ kpc \citep[see][for a recent review]{Sanchez2020}.
Indeed, many of the physical processes that shape global galaxy properties, such as star formation, stellar feedback, and metal redistribution are local processes.
Moreover, up to this point, most large-box galaxy evolutionary simulation models are assessed solely on their ability to reproduce global galaxy properties \citep{Vogelsberger2013,Torrey2014,Schaye2015,Donnari2019}.
Yet, the important physics underpinning these models occurs on sub-galactic scales.
Spatially resolved galaxy scaling relations can provide  diagnostics for galaxies' star formation \citep{Wuyts2013,Hemmati2014,Ellison2020,Hani2020,Pessa2021,Baker2022}, metallicity evolution \citep{Ortega2012,Sanchez2013,Sanchez2019,Patricio2019,Hemler_2021,Garcia_2023,Garcia_2025}, and formation history \citep{Liu2018,Moreno2021,Nelson2021}.
It is therefore critically important to understand galaxies on not just global scales, but local as well.
Quantifying the resolved scaling relations, both in observations and simulation models, is a frontier in our understanding of galaxy evolution.

One scaling relation of interest is the star formation main sequence (SFMS): a tight power-law relation between galactic star formation rate (SFR) and stellar mass ($\mathrm{M}_{\star}$) in normal star forming disk galaxies such that galaxies with larger stellar masses tend to have higher SFRs. 
The global SFMS relation has been observed in surveys at various redshifts~\citep{Brinchman2004,Noeske2007,Daddi2007,Elbaz2007,Leslie_2020,Popesso_2023,Koprowski_2024} as well as reasonably well-reproduced by large-scale cosmological simulations \citep{Dave2011,Torrey2014,Furlong_2015,Sparre2015}.
The overall normalization and power-law index of the SFMS are widely debated in the literature.
Generally speaking, though, the power-law index falls between 0.7-1.3 \citep[e.g.,][]{Noeske2007,Lee_2015,Tomczak_2016}, but the normalization can vary significantly depending on chosen SFR indicator, sample selection, and/or assumed initial mass function \citep{Speagle2014}.

IFS surveys have now measured a {\it resolved} star formation main sequence (rSFMS) on kpc-scales of galaxies \citep[e.g.,][]{Wuyts2013,Hemmati2014,Maragkoudakis2017, Ellison_2018,Ellison2020,Ab2019}.
The rSFMS is a power-law relation between the star-formation rate surface density ($\Sigma_{\rm SFR}$) and stellar mass surface density ($\Sigma_{\star}$).
The power-law index of the rSFMS is qualitatively similar to that of the global SFMS, with surveys reporting values ranging from 0.7-1.1 \citep{Diaz2016, Hsieh2017, Medling_2018, Morselli_2020, Ellison_2021,Pessa2021}.
Interestingly, the rSFMS appears to be relatively invariant with respect to host mass (\citeauthor{Erroz_Ferrer_2019} \citeyear{Erroz_Ferrer_2019}; \citeauthor{Enia_2020} \citeyear{Enia_2020}, although the detailed morphology can change this picture; see, e.g., \citeauthor{Maragkoudakis2017} \citeyear{Maragkoudakis2017}; \citeauthor{Medling_2018} \citeyear{Medling_2018}).
Other factors such as active galactic nuclei (AGN) activity (\citeauthor{Diaz2016} \citeyear{Diaz2016} \citeyear{Diaz_2019}, \citeauthor{Sanchez2018} \citeyear{Sanchez2018}), galaxy-galaxy interactions \citep{Pan_2019,Thorp_2019,Thorp_2022,Brown_T_2023}, and/or environment \citep[][Green et al. In Preparation]{Medling_2018,Vulcani_2020,Sanchez_Garcia_2022,Brown_W_2023} can play a significant role in shaping the star formation activity for individual systems.


Another scaling relation used as a benchmark for galaxy evolutionary models is the relationship between a galaxy's stellar mass and gas-phase metallicity (henceforth mass-metallicity relation, or MZR; see, e.g., \citeauthor{Tremonti2004} \citeyear{Tremonti2004}, \citeauthor{Lee_2006} \citeyear{Lee_2006}, \citeauthor{Kewley2008} \citeyear{Kewley2008} for observational perspective and, e.g., \citeauthor{Dave2011} \citeyear{Dave2011}, \citeauthor{Torrey2017} \citeyear{Torrey2017}, \citeauthor{Garcia_2024c} \citeyear{Garcia_2024c} \citeyear{Garcia_2024b} for theory perspective).
Similar to the SFMS, the MZR provides insight into the star formation history; however, unlike the SFMS, the MZR essentially contains a ``fossil record'' of all previous enrichment and gas mixing within the galaxy.
Heavy metals are formed within stars and then ejected into the interstellar medium (ISM) via supernova explosions and asymptotic giant branch winds.
Once in the ISM, these metals mix via galactic winds and turbulence \citep{Lacey_Fall_1985,Koeppen_1994,Elmegreen_1999} and are diluted by pristine gas inflows from the circum- and inter-galactic media \citep{Somerville_Dave_2015}.
The overall metal content, or metallicity, of a galaxy is therefore an imprint of the processes driving the formation and evolution of galaxies.

The MZR also shares some qualitative similiarities to the SFMS.
The MZR is also a positive power-law correlation for low mass galaxies ($\mathrm{M}_{\star}\,<\,10^{10.0}\,\mathrm{M}_{\odot}$) which flattens, or even turns over, for high mass galaxies ($\mathrm{M}_{\star}\,>\,10^{10.0}\,\mathrm{M}_{\odot}$; see, e.g., \citeauthor{Berg_2012} \citeyear{Berg_2012}, \citeauthor{Andrews_Martini_2013} \citeyear{Andrews_Martini_2013}, \citeauthor{Blanc_2019} \citeyear{Blanc_2019}, \citeauthor{Revalski_2024} \citeyear{Revalski_2024}). 

Analogous to the rSFMS, recent IFS surveys have measured a {\it resolved} MZR \citep[rMZR; e.g.,][]{Ortega2012,Sanchez2013,Sanchez2019,Barrera2016,Barrera_Ballesteros_2018,Zhu2017,Gao_2018}.
Generally, the rMZR follows the same qualitative shape as the MZR: a power-law at low stellar mass surface densities that flattens at the highest stellar mass surface densities.
There is some evidence that the rMZR depends on the stellar mass of the host galaxy \citep{Gao_2018,Hwang_2019} although not all surveys agree \citep{Barrera2016}.
Regardless, the same physical arguments are generally true for chemical enrichment on local scales as global scales -- the metallicity depends on the competition of new enrichment from stars, gas redistribution (i.e., inflows/outflows), and pristine gas accretion \citep{Zhu2017}.

It is generally agreed upon that the resolved scaling relations have the potential to constrain the physics of feedback-regulated star formation on sub-galactic scales given their dependencies on the small-scale physics of the ISM (see, e.g., \citeauthor{Gurvich2020} \citeyear{Gurvich2020}, \citeauthor{Li_2020} \citeyear{Li_2020}, \citeauthor{Nobels_2024} \citeyear{Nobels_2024}).
In this paper, we extend previous simulations results of star-formation and metals on kpc-scales \citep[see][]{Trayford2019,Orr2019,Hani2020,Motwani_2022,McDonough_2023} to the IllustrisTNG cosmological simulation.
We also investigate the dependence (or lack thereof) of the relations on the integrated properties of the host galaxies. We quantify the tightness of the two scaling relations and decouple the scatter of the relation into intrinsic and systematic sources.
Finally, we make a theoretical argument for the emergence of the rSFMS based on the leaky-box model~\citep{Zhu2017}. 

This manuscript is organised as follows. In $\S$\ref{sec:method}, we review the details of our simulation, methods for generating resolved maps for galaxies, and selection criteria for galaxies and spatially resolved regions.
In $\S$\ref{sec:results}, we present our main results for the two scaling relations, their comparison to observations and peer cosmological simulations, the dependence of the relations on host galaxy properties, and the scatters on the relations as intrinsic and systematic sources.
In $\S$\ref{sec:results_rsfms}, we give a discussion about the rSFMS, as a result of the Kennicut-Schmidt relation and the equation-of-state applied in IllustrisTNG models. 
In $\S$\ref{sec:generalized_leaky_box}, we present an extension of the {\cite{Zhu2017}} leaky-box model to account for the rSFMS and $\Sigma_{\rm gas}-\Sigma_\star$ relations.
In $\S$\ref{sec:summary}, we summarize and conclude our results, as well as portray the future direction for our project.

\begin{figure*}
    \centering
    \ifGarcia
        \ifRobinson
            \includegraphics[width=\textwidth]{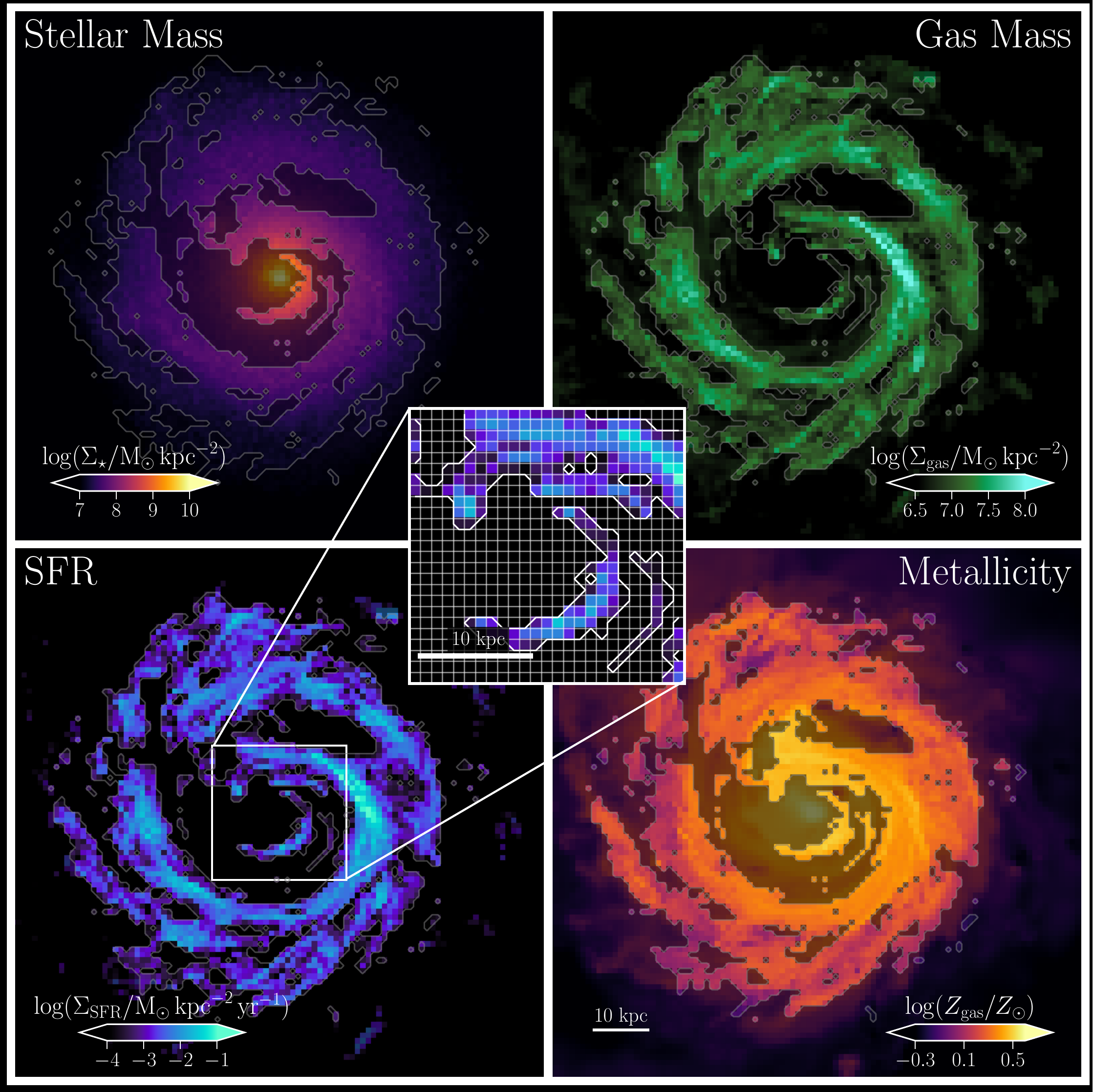}
        \else
            \includegraphics[width=\textwidth]{kpc_maps.pdf}
        \fi
    \else
    \includegraphics[width=\textwidth]{fig2.png}
    \fi
    \caption{{\bf Example of Our Spaxel Construction in IllustrisTNG.} 
    Example face-on map from the TNG50-1 simulation of a galaxy at 1-kpc spaxel resolution.
    \edit{The system shown here (TNG subhalo id $167393$) is a massive star forming galaxy with $\log(M_\star~[M_\odot])=11.66$, $\log(M_{\rm gas}~[M_\odot])=11.62$, and ${\rm SFR}=10.31~M_\odot\,{\rm yr}^{-1}$.}
    We present maps of the stellar mass surface density ($\Sigma_{\star}$; upper left), gas mass surface density ($\Sigma_{\rm gas}$; upper right), star formation rate surface density ($\Sigma_{\rm SFR}$; lower left), and mass-weighted gas-phase metallicity (lower right).
    The field-of-view for the large maps are 100 kpc and each spaxel has a size of 1 kpc $\times$ 1 kpc.
    The central panel shows a zoom-in of the central region of the galaxy in SFR surface density projection.
    \edit{The contours in each panel (loosely) reflect our selection criteria of pixels with $\Sigma_{\rm SFR}>10^{-4}~{M_\odot\,{\rm kpc}^{-2}\,{\rm yr}^{-1}}$ and $\Sigma_\star>10^7~{M_\odot\,{\rm kpc}^{-2}}$ (see Section~\ref{sec:method_map} and Appendix~\ref{appendix:cuts}).}
    We note that these images are down-sampled from the native TNG resolution to be closer to the resolution of current IFS surveys.
    }
    \label{fig:doodle}
\end{figure*}

\section{Methods}
\label{sec:method}

\subsection{IllustrisTNG}\label{sec:method_simulation}

In this paper, we employ data products from {The Next Generation} (TNG) of Illustris simulations~\citep{Pillepich2018b, Springel2018, Marinacci2018, Naiman2018, Nelson2018, Nelson_2019}.
TNG is a series of hydrodynamic cosmological simulations of galaxy formation within the $\Lambda$CDM paradigm. 
TNG \edit{models many} physical processes governing galaxy formation: gravity, hydrodynamics, star formation, stellar evolution, chemical enrichment, primordial and metal-line cooling of the gas, stellar feedback with galactic outflows, and black hole formation, growth and feedback \citep[see][for a complete description of the TNG model]{Pillepich2018a}. 

The TNG simulations include three box sizes, each simulated at three resolution levels~\citep{Pillepich2018b}.
The analysis in this work centers on the highest resolution run of the $(35h^{-1}~ \mathrm{Mpc})^3$ box: {TNG50-1} (hereafter simply TNG; \citeauthor{Nelson_2019b} \citeyear{Nelson_2019b}, \citeauthor{Pillepich_2019} \citeyear{Pillepich_2019}).
This simulation contains a total of $2\times2160^3$ particles and a baryon mass resolution of $8.5\times10^4M_\odot$.

\edit{
It is important to recognize that the sites of star formation (giant molecular clouds) are unresolved in TNG.
Many of the aforementioned physical processes are thereby also unresolved in the simulation and are necessarily implemented on ``sub-grid'' scales.
The TNG model provides a physically-motivated (and numerically necessary) prescription for the behavior of the ISM, stars, and black holes on these unresolved scales.
}
\edit{As an example, }the TNG model treats the star-forming ISM with the \cite{Springel2003} effective equation of state.
The effective equation of state allows new star particles to form stochastically in dense ($n_{\rm H} > 0.13~{\rm cm}^{-3}$) ISM consistent with a \cite{Chabrier2003} initial mass function.
\edit{This dense gas in TNG forms stars based on a {\it volumetric} version of a Schmidt-Kennicutt relation, relating the 3D SFR density ($\dot{\rho}_\star$) to the 3D gas density ($\dot{\rho}_{\rm gas}$) via
\begin{equation}
    \dot{\rho}_\star = \frac{\rho_{\rm gas}}{\tau_{\rm ff}}~,
\end{equation}
where $\tau_{\rm ff}$ is the free fall time, which itself is assumed to be proportional to $\rho_{\rm gas}^{-0.5}$.
Star formation therefore takes place such that
\begin{equation}
    \label{eqn:volumetric_SK}
    \dot{\rho}_\star \propto \rho_{\rm gas}^a
\end{equation}
where $a$ is $1.5$.
The existence of the (resolved) Schmidt-Kennicutt relation in these models is therefore not emergent, it is prescribed (exactly by Equation~\ref{eqn:volumetric_SK}, where the normalization is a tuned parameter of the model).
}

At the end of the lives of these stars (the time-scales of which are set by \citeauthor{Portinari_1998} \citeyear{Portinari_1998} and depend on the mass and metallicity of the star), both mass and metals are returned into the ISM through either asymptotic giant branch (AGB) winds or Type II supernovae.
The elemental yields of these events follow prescriptions from \cite{Nomoto_1997}, \cite{Portinari_1998}, \cite{Kobayashi_2006}, \cite{Karakas_2010}, \cite{Doherty_2014}, and \cite{Fishlock_2014}.
TNG explicitly tracks \edit{the evolution of} 9 chemical species (H, He, C, N, O, Ne, Mg, Si, and Fe) as well as a tenth ``other metals'' field which is a proxy for the metal species not explicitly tracked.

The TNG model also includes contributions from active galactic nuclei (AGN) in addition to stellar evolution and feedback \citep[see][for a complete review of the TNG AGN model]{Weinberger_2017}.
\edit{Supermassive black holes (SMBHs) of mass $1.18\times10^6M_\odot$ are seeded -- i.e., manually placed -- at location of the potential minimum in halos once their total mass exceeds $7.38\times10^{10}M_\odot$.
The SMBHs then accrete gas according to a Eddington-limited Bondi accretion rate.
}
TNG implements black hole-driven feedback in two modes: \edit{``radio'' and ``quasar'' accretion.
The mode of feedback a SMBH uses is set by the accretion rates with the transition occurring at a threshold of 
\begin{equation}
    \chi = \operatorname{min}\left[0.002\left(\frac{M_{\rm BH}}{10^8 M_\odot}\right)^2,\,0.1\right]~,
\end{equation}
where $\chi$ is the accretion rate (in units of Eddington accretion) and $M_{\rm BH}$ is the mass of the SMBH.}
The high-accretion rate mode of AGN feedback (quasar mode) provides thermal energy into the surrounding gas \edit{with a rate of $\dot{E}_{\rm therm}=0.02\dot{M}c^2$.}
The low-accretion rate mode (radio mode) implements \edit{pulsed, directed} kinetic feedback in the form of black hole driven winds \edit{with rate $\dot{E}_{\rm kin} = \epsilon_{f,\,{\rm kin}}\dot{M}c^2$ (where $\epsilon_{f,\,{\rm kin}}$ depends on the local density relative to the star formation rate density threshold)}.
The net result of the AGN feedback within the TNG simulations is to significantly reduce the star formation in the most massive halos \citep{Weinberger_2017}.

\subsection{Galaxy Identification and Sample Selection}\label{sec:method_selection}

Dark matter haloes are identified using a friends-of-friends (FOF) algorithm \citep{Davis1985} in TNG, from which individual galaxies are identified using the {\sc subfind} algorithm \citep{Springel2001}.
From these {\sc subfind}-identified galaxies, we impose \edit{two} general selection criteria.
First, we require all simulated galaxies to have 
\edit{a large enough sample of particles to robustly determine its spatially resolved properties.
In testing, we find this to be $\gtrsim10^4$ particles -- corresponding to galaxies with $M_\star > 10^9\,\mathrm{M}_\odot$ (given baryon mass resolution of $m_{\rm baryon}=8.5\times10^4M_\odot$).}
Secondly, we use a star formation rate threshold of \edit{$\mathrm{SFR} > 10^{-2}\,M_\odot\,\mathrm{yr}^{-1}$} to select star-forming galaxies.
The second selection criterion also ensures that there are star-forming gas cells {\it somewhere} within the system.
This is desirable as observational metallicity measurements are strictly limited emission from bright star forming regions in galaxies \citep{Kewley2008,Kewley_2019,Maiolino_Mannucci_2019}.
\edit{
We note that our selection criteria is agnostic to whether a galaxy is the ``central'' (i.e., most massive in its group); however, we find that the scaling relations of central and satellite systems do not vary significantly.
}
Taken together, these selection criteria yield \edit{2,734}
galaxies at $z=0$.




\subsection{Resolved spaxel maps}\label{sec:method_map}

A central goal of this work is to compare the TNG simulated resolved galaxy scaling relations against observations -- specifically those of CALIFA~\citep{Sanchez2012}, MaNGA~\citep{Bundy2015} and PHANGS \citep{Leroy_2021}. 
Such campaigns investigate the rSFMS and gas-phase rMZR with the resolution of about $1$ kpc at low redshift ($z<0.1$; \citeauthor{Maragkoudakis2017} \citeyear{Maragkoudakis2017}).
We note that (at $z\,=\,0$) the smallest hydrodynamical gas cells in TNG have an extent of $48\,\mathrm{pc}$ and most baryonic cells are smaller than $1\,\mathrm{kpc}$ -- slightly higher than these current IFS studies.
Our spaxel map construction choices, therefore, do not represent the native spatial resolution of TNG; instead, they are adjusted to be more comparable to the resolution of observations.

We create two-dimensional maps of our galaxies for each relevant galaxy property for this study: the stellar mass surface density ($\Sigma_{\star}$), gas mass surface density ($\Sigma_{\rm gas}$), star formation rate surface density ($\Sigma_{\rm SFR}$), and gas phase metallicity (each illustrated in Figure~\ref{fig:doodle}).
The pixel (spaxel) size in these images (and, indeed, in the entirety of our analysis) is set to a fixed value of 1 kpc on the side. 
To construct these spaxel maps, we extract particles associated with an individual galaxy based on the FOF halo assignments. 
We then center the galaxy based on its center-of-mass.
Next, we rotate the galaxy to a face-on orientation by calculating the net angular momentum vector for all gas particles within $5$ kpc of the galactic center and aligning that angular momentum vector to the $\hat z$ direction.
This alignment maximizes the number of valid spaxels within each galaxy and promotes a fair comparison with the observed datasets, which preferentially select low-inclination systems\ignorespaces
\footnote{\ignorespaces
Recently, EDGE-CALIFA (\citeauthor{Bolatto_2017} \citeyear{Bolatto_2017}) and GECKOS (\citeauthor{vandeSande_2024} \citeyear{vandeSande_2024}) have begun to include edge-on or highly inclined systems.
While interesting comparisons exist between these edge-on surveys, we opt to primarily focus on face-on systems and ignore inclination effects in this work.
}. 
We note that our method for disk alignment is imperfect for a subset of systems, especially those with warped disks; however, this method efficiently brings the majority of systems into face-on alignment.
Moreover, the small residual angular offsets away from perfect face-on orientation in these warped systems would not significantly affect our main results.
Finally, we create a 3D grid for each property of interest (i.e., stellar mass, gas mass, SFR, and gas-mass-weighted gas metallicity) by mapping the discrete particle distributions onto a 3D grid using a traditional smoothing method with a cubic spline kernel.
This 3D grid is then converted into a 2D map by integrating over the $\hat{z}$ direction for each 2D spaxel.
We \edit{find that reasonable variations to this procedure do not qualitatively change our core results (see Appendix~\ref{appendix:density_reconstruction}, which is in agreement with \citeauthor{Trayford2019} \citeyear{Trayford2019} their Appendix B)}. 

\edit{With the maps created, we make two additional cuts on our data (shown visually in Figure~\ref{fig:doodle}).}
We \edit{first} require that individual spaxels to contain stellar mass surface densities of $\Sigma_{\star}\,>\,10^7\,\mathrm{M}_{\odot}\,\mathrm{kpc}^{-2}$.
\edit{Given our pixel size of $1~{\rm kpc}\times 1~{\rm kpc}$, the stellar mass surface density cut is made to ensure that each pixel is well-sampled by the simulation particles.
The cut of $\Sigma_{\star}\,=\,10^7\,\mathrm{M}_{\odot}\,\mathrm{kpc}^{-2}$ roughly corresponds to the contribution of $100$ star particles per individual pixel given the mass resolution of TNG50.
}
The second cut is a star formation rate surface densities cut of $\Sigma_{\rm SFR}\,>\,10^{-4}\,\mathrm{M}_{\odot}\,\mathrm{yr}^{-1}\,\mathrm{kpc}^{-2}$ (which implicitly also requires the presence of gas mass in a pixel, \edit{see Equation~\ref{eqn:volumetric_SK}}).
\edit{This second cut is to make our analysis more analogous to current IFU studies, which have detection limits of a similar order of magnitude \citep{Diaz2016,Hsieh2017}.
Moreover, this star formation surface density cut ensures that each pixel is star forming given our stellar mass surface density cut (i.e., has $\Sigma_{\rm sSFR} > 10^{-11}~{\rm yr^{-1}kpc^{-2}}$).
}
\edit{We note that other reasonable, less restrictive cuts do not significantly impact our core results (Appendix~\ref{appendix:cuts}). }
From the selected galaxies \edit{and pixel selection criteria,} we end up with \edit{375,033}
star-forming spaxels that are used in our analysis.

\section{Results}\label{sec:results}

\subsection{The Resolved Star Formation Main Sequence (rSFMS)}\label{sec:results_rsfms}

\begin{figure}
    \centering
    \ifRobinson
        \includegraphics[width=\columnwidth]{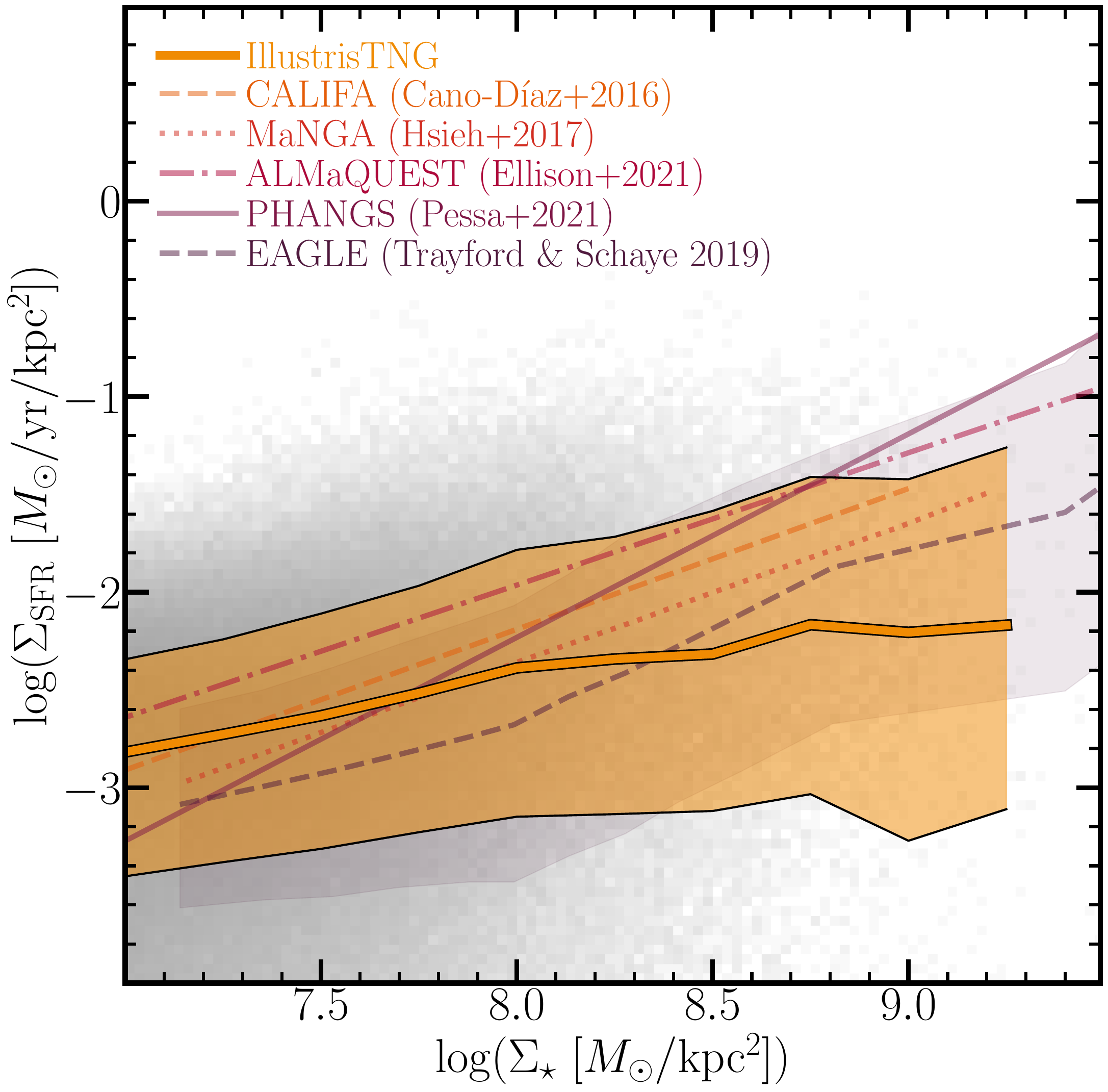}
    \else
        \includegraphics[width=\columnwidth]{hist2d_sfr.pdf}
    \fi
    \caption{{\bf The Resolved Star Formation Main Sequence (rSFMS) in IllustrisTNG}
    The rSFMS in TNG at $z\,=\,0$ with 1 kpc spaxel resolution.
    The solid orange line shows the median value of $\mathrm{\Sigma}_{\mathrm{SFR}}$ in each $\Sigma_{\star}$ bin, whilst the two dashed orange lines indicate the 16th-84th percentiles.
    The background 2D distribution displays our selected spaxels, color-coded by the number in each bin. 
    For comparison, we show the rSFMS from the EAGLE simulation \protect\citep{Trayford2019} in the dashed line (and shaded region about it), observational results from the MaNGA survey \protect\citep{Hsieh2017} in the dotted line, from the CALIFA survey \protect\citep{Diaz2016} in the red dashed line, and PHANGS survey in the solid black line \protect\citep{Pessa2021}, and ALMaQUEST \protect\citep{Ellison_2021} in the dot-dashed line.
    We note that for the rSFMS, as well as all other median/percentile binning in this work, we space $\Sigma_\star$ bins in width of 0.25 dex and require 200 spaxels in each bin.
    }
    \label{rsfms}
\end{figure}

We present the $z=0$ rSFMS from TNG in Figure~\ref{rsfms}.
The 2D histogram shows the distribution of all valid spaxels from the TNG galaxy population, with darker histogram pixels indicating more spaxels in that bin.
The thick solid orange line represents the median of the distribution, with the shaded orange region representing the 16-84$^{\rm th}$ percentile of the distribution.
Typically the rSFMS is fit with a power-law described by
\begin{equation}
    \label{eqn:rSFMS}
    \edit{
    \Sigma_{\rm SFR} = \beta\Sigma_{\star}^\alpha,
    }
\end{equation}
where $\alpha$ is the power-law index (i.e., slope in log-log space) and $\beta$ is the scaling coefficient.
Indeed, the median relation in TNG is well-described by this power-law relation with an index of \edit{$\alpha = 0.302$} in the range $\Sigma_* = 10^7 - 10^{9} \, \mathrm{M}_\odot \, \mathrm{kpc^{-2}}$.
We also include the median (or best-fit, in some cases) rSFMS from recent IFS surveys: CALIFA as the dashed orange line \citep[from][]{Diaz2016}, MaNGA as the dotted red line \citep[from][]{Hsieh2017}, ALMaQUEST in the dot-dashed red line \citep[from][]{Ellison_2021}, and PHANGS as the solid purple line \citep[from][]{Pessa2021}.
The power-law indeces from these observational surveys \edit{tend to be steeper than} that of TNG \citep[see also][]{Medling_2018,Morselli_2020}\ignorespaces
\footnote{\ignorespaces
We do note, however, that different fitting procedures can change the slope significantly (see, e.g., \citeauthor{Lin_2019} \citeyear{Lin_2019}; \citeauthor{Morselli_2020} \citeyear{Morselli_2020}; \citeauthor{Ellison_2021} \citeyear{Ellison_2021}).
}.
\edit{In the following sections, we argue that the disagreement between TNG and simulations can be largely attributed to the host galaxy properties.}

The normalization of the rSFMS, \edit{on the other hand}, is reasonably well reproduced by TNG compared to observational surveys.
The broad agreement in normalization of the rSFMS is notable as the TNG physical model is not tuned in any direct way to give rise to the observed rSFMS. 
Rather, this simulated rSFMS is a consequence of the underlying galaxy formation model already included in TNG \citep{Springel2003}.
We also include predictions from the EAGLE simulation \citep[dashed black line;][]{Trayford2019}.
We find \edit{reasonable} agreement between the two different simulation models in terms of \edit{normalization} and scatter; however, \edit{just as in observations,} the \edit{power-law index} of the rSFMS in EAGLE is \edit{steeper} than in TNG.

\edit{
The scatter about the rSFMS is $\sim0.5$ dex, which is larger than the rSFMS of observations \citep[see, e.g.,][]{Diaz_2019}.
}
\edit{Moreover,} the full population of data points about the rSFMS occupy \edit{a very broad range of star formation rate surface densities at fixed stellar mass surface density (see, e.g., Figure~\ref{fig:resolved_pixel} in Appendix~\ref{appendix:cuts}).
In particular, the range of $\Sigma_{\rm SFR}$, when removing our valid pixel criteria, includes pixels with contributions at $\lesssim10^{-6} M_\odot/{\rm yr}/{\rm kpc}^2$.
We note that the existence of such extremely low star formation rate pixels is likely, in part, related to our parameter reconstruction method (which we investigate further in Appendix~\ref{appendix:cuts}).
Moreover, equivalent regions in observational studies would be impossible to detect given the sensitivity of current IFU instruments.
We therefore caution against a strong interpretation of these very low $\Sigma_{\rm SFR}$ pixels.
}
\edit{Regardless, the inclusion of these lower $\Sigma_{\rm SFR}$ pixels does not significantly impact the core results of this work.}

\edit{To understand what drives the scatter in pixels which {\it do} pass our selection criteria,} we investigate the dependence on host mass within the TNG model in the following sub-section.

\subsubsection{Dependence of rSFMS on Host Galaxy Mass}
\label{subsubsec:rSFMS_mass_TNG}

\begin{figure*}
    \centering
    \ifGarcia
        \ifRobinson
            \includegraphics[width=\linewidth]{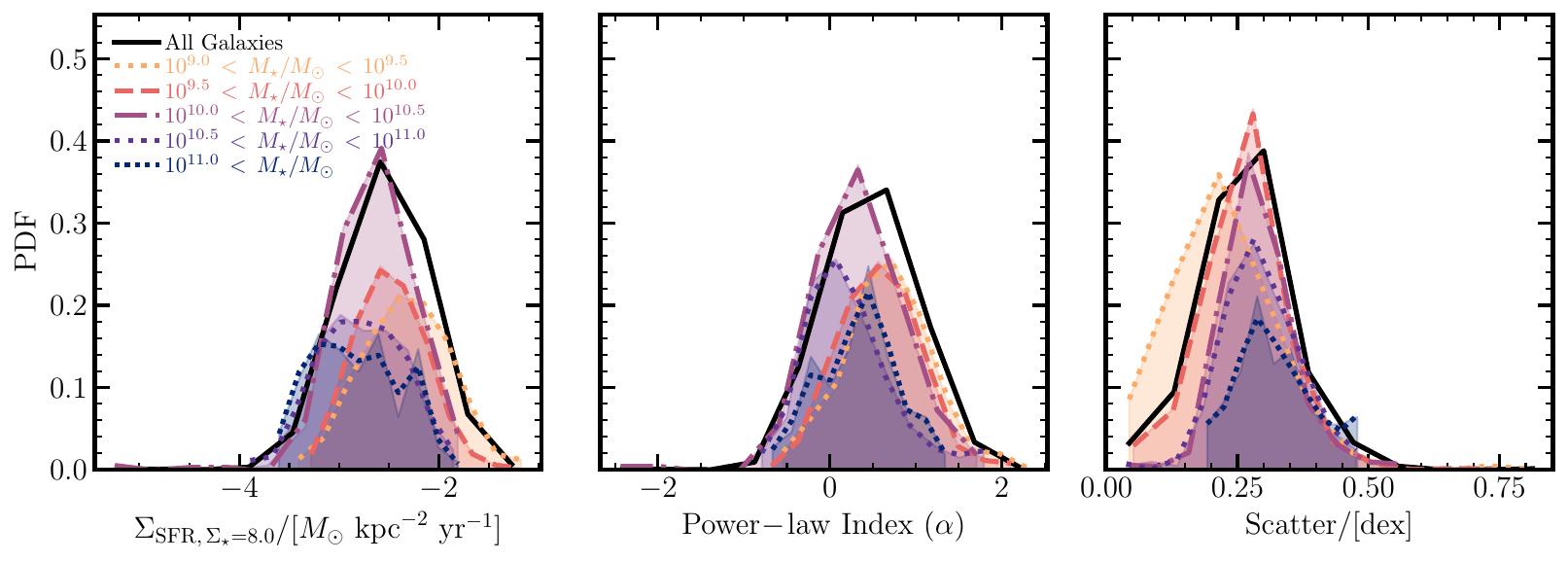}
        \else
            \includegraphics[width=\textwidth]{Figure2.pdf}
        \fi
    \else
        \includegraphics[width=0.33\textwidth]{all_norms.pdf}
        \includegraphics[width=0.33\textwidth]{all_scatters.pdf}
        \includegraphics[width=0.33\textwidth]{all_slopes.pdf}
    \fi
    
    \caption{{\bf Best-fitting parameters for the rSFMS of individual galaxies.}
    {\it Left:} The probability distribution function (PDF) of normalizations of the rSFMS for individual galaxies $\Sigma_{\mathrm{SFR}}$ (taken at $\mathrm{\Sigma}_{\star}\,=\,10^{8.0}\,\mathrm{M}_{\odot}$). 
    In all three panels the colored distributions  are the individual galaxies' PDFs broken up by global mass bin and the solid black line is the distribution for all galaxies combined.
    {\it Middle:} The PDF of power-law indexes (i.e., $\alpha$ from Equation~\ref{eqn:rSFMS}).
    {\it Right:} The scatter distribution around their best-fit linear rSFMS model, as measured by the standard deviation of residual. }
    \label{fit_all}
\end{figure*}

First, we quantify the effect of individual systems by extracting spaxel data for systems with more than \edit{50}
valid spaxels (see Section.~\ref{sec:method_selection} for definition of ``valid'') and fit the rSFMS for each individual galaxy\ignorespaces
\footnote{\ignorespaces
The choice of \edit{50} valid spaxels is to balance between the meaningful nature of an individual galaxy fit (i.e. fits with very few pixels may easily become noise-dominated) and the total number of galaxies analyzed.
With too few valid spaxels, we cannot fit a meaningful rSFMS for individual spaxels, but 
requiring too many valid spaxels significantly reduces the number of available galaxies.
We note that our key results are not sensitive to this choice of spaxel counts. 
}. 
For brevity, we summarize the results of this analysis with three parameters describing the rSFMS within individual galaxies: 
(i) $\mathrm{\Sigma}_{\mathrm{SFR}}$ at $\mathrm{\Sigma}_{\star}\,=\,10^{8.0}\,\mathrm{M}_{\odot}\,\mathrm{kpc}^{-2}$,
(ii) the best-fitting power-law indexes ($\alpha$ from Equation~\ref{eqn:rSFMS}) for individual galaxies, and 
(iii) the standard deviation of spaxel's $\Sigma_{\rm SFR}$ around the regression.
Figure~\ref{fit_all} shows the distribution of the parameters of individual rSFMS for all galaxies (black line) as well as galaxies within different stellar mass bins (colored lines).

The $\Sigma_{\mathrm{SFR}}$ of most systems ranges between \edit{$\sim10^{-3.5}$} to $\sim10^{-2.0}\,\mathrm{M}_{\odot}\,\mathrm{kpc}^{-2}\,\mathrm{yr}^{-1}$ at $\Sigma_{\star}\,=\,10^{8.0}\,\mathrm{M}_{\odot}\,\mathrm{kpc}^{-2}$, with a median of \edit{$10^{-2.53}\,{M}_{\odot}\,\mathrm{kpc}^{-2}\,\mathrm{yr}^{-1}$ for all galaxies}. 
This median normalization is broadly consistent with the median relation in Figure~\ref{rsfms}.
Lower mass galaxies tend to have higher $\Sigma_{\mathrm{SFR}}$ \edit{(median of $10^{-2.38}M_\odot\,{\rm yr}^{-1}\,{\rm kpc}^{-2}$)} whereas highest mass galaxies tend to have lower $\Sigma_{\mathrm{SFR}}$ \edit{(median of $10^{-2.83}M_\odot\,{\rm yr}^{-1}\,{\rm kpc}^{-2}$)} at $\Sigma_{\star}\,=\,10^{8.0}\,\mathrm{M}_{\odot}\,\mathrm{kpc}^{-2}$. 
\edit{It should be noted, the difference offset in median normalization ($\sim0.5$ dex) is on the same order as that of the scatter about the combined rSFMS.}
The right panel of Figure~\ref{fit_all} shows the intrinsic pixel-to-pixel variations within the same galaxies.
For all galaxies, the $1\sigma$ scatter is centered at $\sim0.25$ dex, which is \edit{similar to the combined rSFMS} \edit{and that of the observed rSFMS \citep[][]{Diaz_2019,Ellison_2021}}.
\edit{In the individual mass bins, the median scatter seems to increase monotonically with increasing host mass, from $0.21$ dex in the lowest mass bin to $0.32$ dex in the highest mass bin.}

The median power-law index for all individual galaxies is \edit{$0.48$}, which is \edit{steeper than} the slope of rSFMS by stacking all spaxels together (\edit{$0.3$}).
Interestingly, the rSFMS is \edit{significantly} and shallower for higher-mass galaxies: \edit{with the lowest mass galaxies ($10^{9.0}M_\odot < M_\star < 10^{9.5}M_\odot$) having an median power-law index of $0.66$ and the highest mass galaxies ($M_\star > 10^{11.0}$) of $0.3$.}
In fact, a significant fraction of galaxies even show an anti-correlation between $\Sigma_{\star}$ and $\Sigma_{\mathrm{SFR}}$ in \edit{mass bins ($M_\star > 10^{10.5}M_\odot$)}.
These inversions are, in all likelihood, caused by AGN feedback, which \edit{primarily} effects massive galaxies in the TNG model.
AGN have been shown to suppress star-formation rates of galaxies within TNG \citep{Weinberger_2017}.
\edit{The power-law indices derived for the highest mass bins are closest to the combined relation while the rSFMS derived for the lower mass systems are systematically closer to the power-law indices of observed galaxies \citep{Diaz_2019}.
Indeed, the bulk of pixels in our sample ($\sim60\%$) come from systems with stellar masses of $>10^{10.5}M_\odot$, in part because the more massive systems are simply larger than their low mass counterparts \citep{Genel_2018}.
The disagreement between the observed and TNG rSFMS is therefore likely driven by the high mass systems. 
}


\begin{figure}
    \ifGarcia
        \centering
        \ifRobinson
            \includegraphics[width=\linewidth]{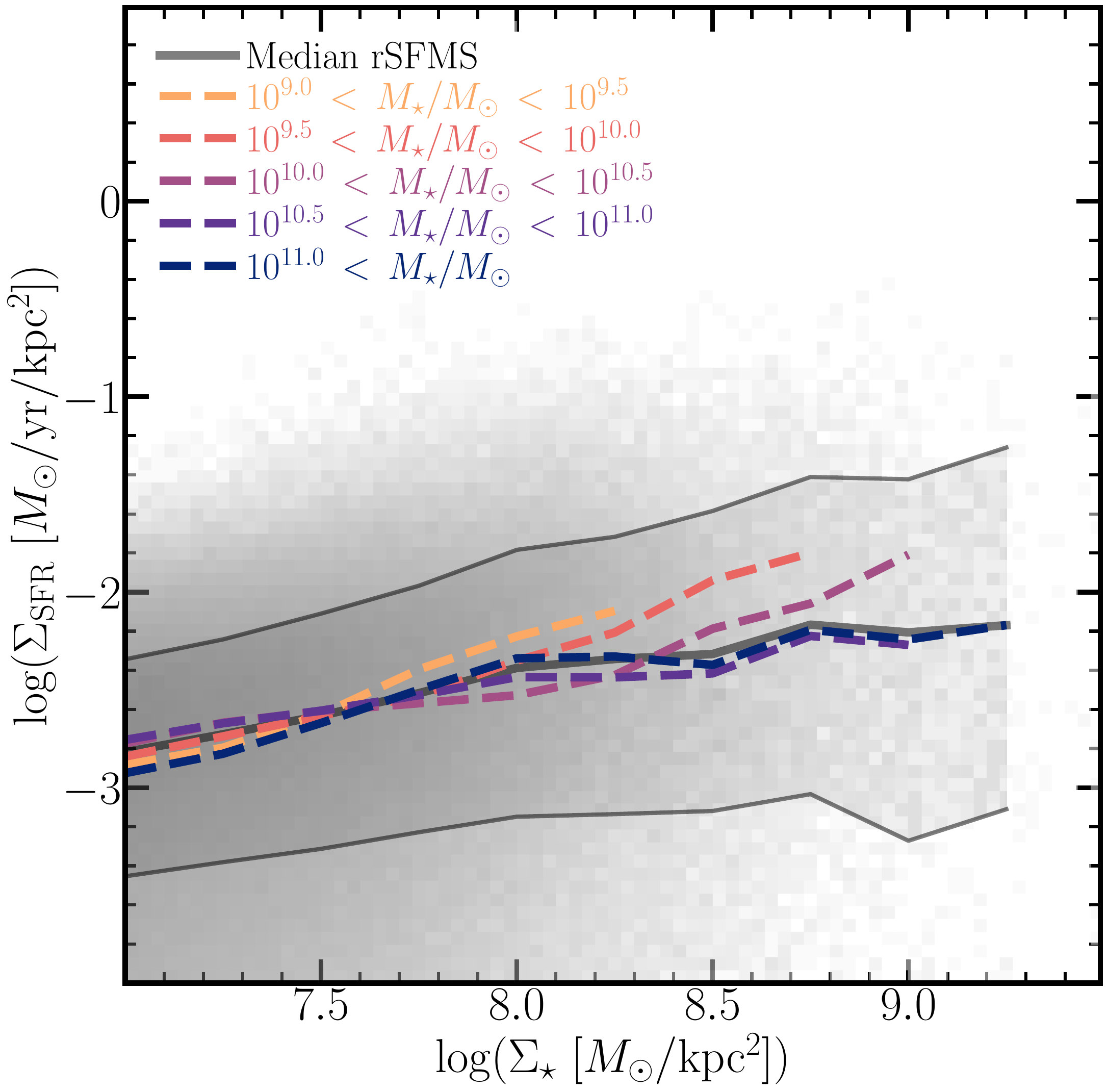}
        \else
            \includegraphics[width=\linewidth]{hist2d_sfr_bins.pdf}
        \fi
    \else
        \centering{
        \includegraphics[width=0.45\textwidth]{hist2d_sfr_mass090.png}
        \includegraphics[width=0.45\textwidth]{hist2d_sfr_mass095.png}}
        \centering{
        \includegraphics[width=0.45\textwidth]{hist2d_sfr_mass100.png}
        \includegraphics[width=0.45\textwidth]{hist2d_sfr_mass105.png}
        }
    \fi
    \caption{
    {\bf The rSFMS for galaxies binned by total mass.}
    The rSFMS for galaxies within different mass bins as indicated by the title. The black solid line plots the median rSFMS relation and the black dashed line indicates the $1\sigma$ scatter. The colored background shows the distribution of spaxels, the same as Figure~\ref{rsfms}. 
    }
    \label{rsfms_com} 
\end{figure}

\begin{table}
    \centering
    \begin{tabular}{lcc}
        \toprule
         \textbf{Mass bin} ($\log M_\odot$) & \textbf{Power-Law} ($\alpha$) & \textbf{Coefficient} ($\beta$) \\\midrule
         All & $0.302$ & $10^{-4.883}$\\\midrule
         $9.0-9.5$   & $0.671$ & $10^{-7.625}$\\
         $9.5-10.0$  & $0.606$ & $10^{-7.157}$\\
         $10.0-10.5$ & $0.445$ & $10^{-5.963}$\\
         $10.5-11.0$ & $0.250$ & $10^{-4.484}$\\
         $11.0+$     & $0.332$ & $10^{-5.151}$ \\\bottomrule
    \end{tabular}
    \caption{{\rm rSFMS fit parameters.} The best-fit power-law index ($\alpha$) and the scaling coefficient ($\beta$) of Equation~\ref{eqn:rSFMS} for each stellar mass bin's rSFMS as well as the total rSFMS.
    }
    \label{tab:rSFMS}
\end{table}

To \edit{further illustrate} the dependence on the spaxels' host, Figure~\ref{rsfms_com} shows the rSFMS for spaxels split into different bins by the stellar mass of their host galaxies.
The normalization of each of the mass-binned rSFMS is similar to each other for stellar mass surface densities below $\Sigma_* < 10^{7.5} \mathrm{M}_\odot \mathrm{kpc}^{-2}$. 
At higher $\Sigma_\star$, however, lower mass systems tend to have higher $\Sigma_{\mathrm{SFR}}$.
To put the differences between the different mass bins more concretely, we fit each mass bin with its own power-law (of the same form as Equation~\ref{eqn:rSFMS}).
We present the power-law index and scaling coefficient for the rSFMS mass bins in Table~\ref{tab:rSFMS}.
We find that with increasing mass bin the power-law index decreases, while the scaling coefficient increases \edit{mirroring the results of the previous section}.

\edit{Overall, we find fairly significant variations in the rSFMS in galaxies of different stellar mass.
The differences here are interesting as there is not such a strong scaling of rSFMS properties seen in observations \citep[e.g.,][]{Diaz_2019}.
This may suggest that the AGN feedback implementation in TNG, which we believe to be responsible for the variations seen at higher stellar masses, may be too significant.
Indeed, \cite{Corcho-Caballero_2023} shows that the AGN feedback implementation in TNG leads to too many quenched galaxies and \cite{Suresh_2025} argue that the AGN feedback of several simulation models (including TNG) are inconsistent with observations.
We explore the mass dependence of the rSFMS further in Section~\ref{subsec:mass_rSFMS}.
}

\subsection{The Resolved Mass Metallicity Relation (rMZR)} 
\label{sec:results_rmzr}

\begin{figure}
    \centering
    \ifGarcia
        \ifRobinson
            \includegraphics[width=\columnwidth]{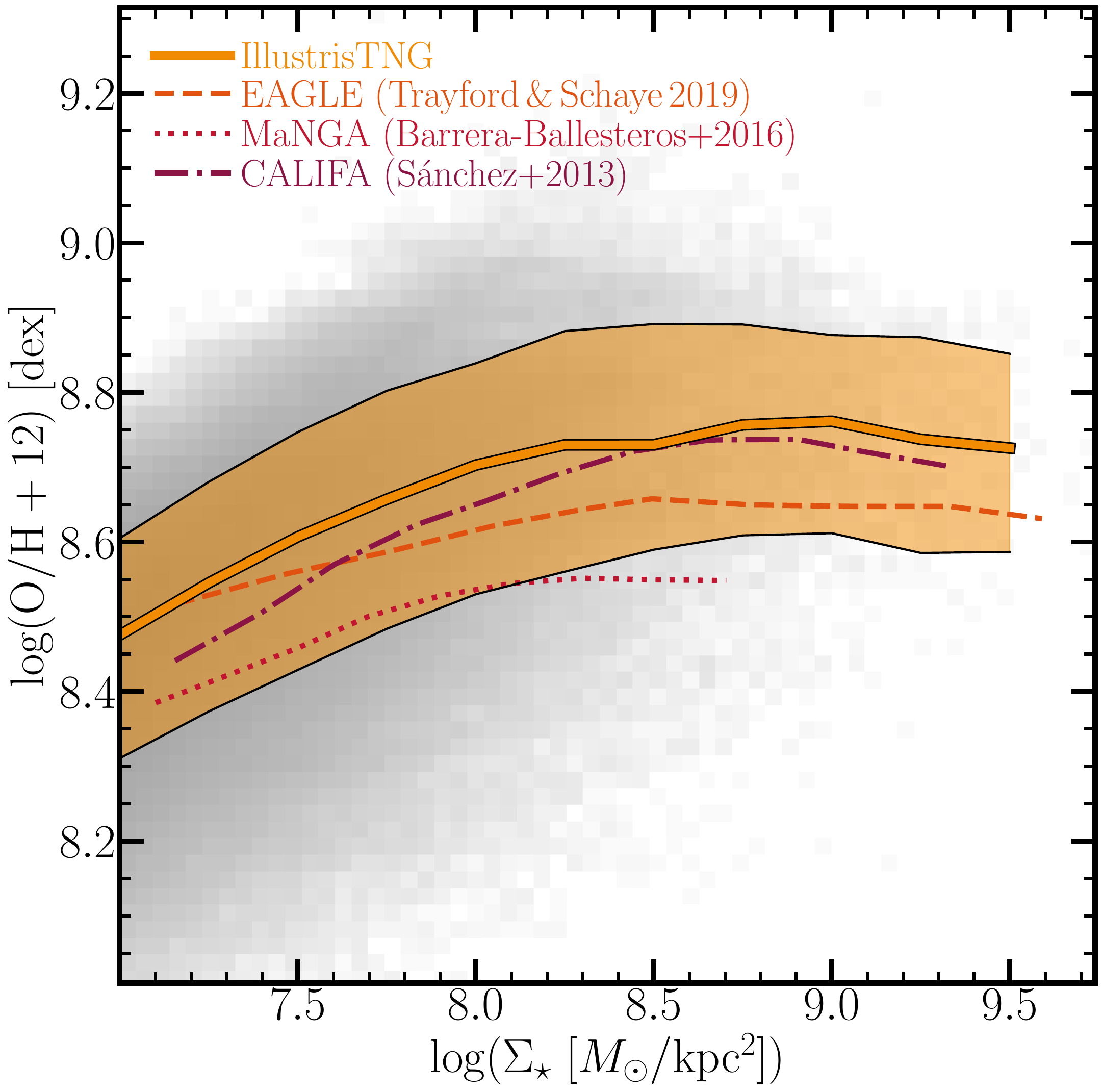}
        \else 
            \includegraphics[width=\columnwidth]    {hist2d_z.pdf}
        \fi
    \else
        \includegraphics[width=\columnwidth]{hist2d_z.pdf}
    \fi
    \caption{
    {\bf The Resolved Mass Metallicity Relation (rMZR) in IllustrisTNG.}
    The rMZR in TNG at $z\,=\,0$ with spaxel resolution of $1\,\mathrm{kpc}$ \edit{(shifted by $-0.2$ dex)}.
    The solid line shows the median value of $12\,+\,\mathrm{log}(\mathrm{O}/\mathrm{H})$ and  the shaded region indicates the 16$^{\rm th}$-84$^{\rm th}$ percentile scatter of the background distribution of spaxels.
    For comparison, the dashed line is the analogous result from the EAGLE simulation (\protect\citeauthor{Trayford2019} \protect\citeyear{Trayford2019}; \edit{which itself is shifted by $-0.6$ dex from the true metallicity values of EAGLE}).
    The dotted line is from the MaNGA survery (\protect\citeauthor{Barrera2016} \protect\citeyear{Barrera2016}) and the dash-dotted line is from the CALIFA survery (\protect\citeauthor{Sanchez2013} \protect\citeyear{Sanchez2013}).
    We caution against too strong a comparison against observational results, however; see text for more details.
    }
    \label{rzms}
\end{figure}

Figure~\ref{rzms} shows the rMZR at $z\,=\,0$ from TNG.
The background 2D histogram shows the full distribution of spaxels, while the solid line indicates the median rMZR relation and the shaded region is $16^{\rm th}-84^{\rm th}$ percentile range.
The rMZR of TNG follows the same general shape as the integrated MZR: a power-law at low stellar mass (surface densities) with a flattening at the highest masses.
More specifically, the median rMZR \edit{in TNG} is well-described by a positive power-law below a surface density $\Sigma_{\star}\,<\,10^{8.5}\,\mathrm{M}_{\odot}\,\mathrm{kpc}^{-2}$ with a power-law index of \edit{$0.21$}.
Above $\Sigma_{\star}\,\sim\,10^{8.5}\,\mathrm{M}_{\odot}\,\mathrm{kpc}^{-2}$, however, the relation flattens significantly and plateaus.
These high-$\Sigma_{\star}$ spaxels tend to inhabit the central regions ($\mathrm{r}\,\lesssim\,2\,\mathrm{kpc}$) of massive galaxies.
The spaxels at small galactocentric distances are therefore likely to be strongly affected by AGN-driven outflows in the TNG model (see Section~\ref{sec:method_simulation} and \citeauthor{Weinberger_2017} \citeyear{Weinberger_2017} for more details on TNG AGN modeling).

Next, we compare against observations -- CALIFA  \citep{Sanchez2013} and MaNGA \citep{Barrera2016} -- as well as other simulation results \citep[EAGLE,][]{Trayford2019}.
We caution against too strong a comparison between the \edit{normalization of the} rMZR in TNG with that of observations as the derived metallicities from the IFS surveys are dependent on the chosen diagnostic \citep{Kewley2008, Kewley_2019,Maiolino_Mannucci_2019}.
\edit{In fact, to avoid a direct interpretation of the normalization of the rMZR, we adjust the normalization of the TNG relation by $-0.2$ dex to match closer to the observed relations (following from \citeauthor{Trayford2019} \citeyear{Trayford2019}, wherein the EAGLE rMZR was adjust by $-0.6$ dex).}
We therefore strongly advocate against a detailed comparison of the normalization of the rMZR without a more careful treatment of the diagnostics.
Regardless, we opt to make the comparison of the {\it qualitative} shape of the rMZR in order to directly compare our results to previous results\ignorespaces
\footnote{\ignorespaces
We caution, however, that the shape of the (r)MZR may also depend on diagnostics (see, e.g., \citeauthor{Kewley2008} \citeyear{Kewley2008}).
}.
We find that the shape of the rMZR in IllustrisTNG broadly matches its observational counterparts and EAGLE predictions.
Their common characteristics are captured in the IllustrisTNG result: as a positive power-law for low $\Sigma_{\star}$ regime and a flattening/turnover for high $\Sigma_{\star}$.
The TNG rMZR is steeper than that of MaNGA and EAGLE for the positively-correlated low $\Sigma_{\star}$ regime, but quite similar to that of CALIFA.
Similarly, the flattening starts at a lower $\Sigma_{\star}$ in EAGLE and MaNGA, but is reasonably similar to CALIFA. 

\subsubsection{Dependence of rMZR on Host Galaxies}
\label{subsubsec:rMZR_host}

\begin{figure}
    \centering
    \ifGarcia
        \ifRobinson
            \includegraphics[width=\linewidth]{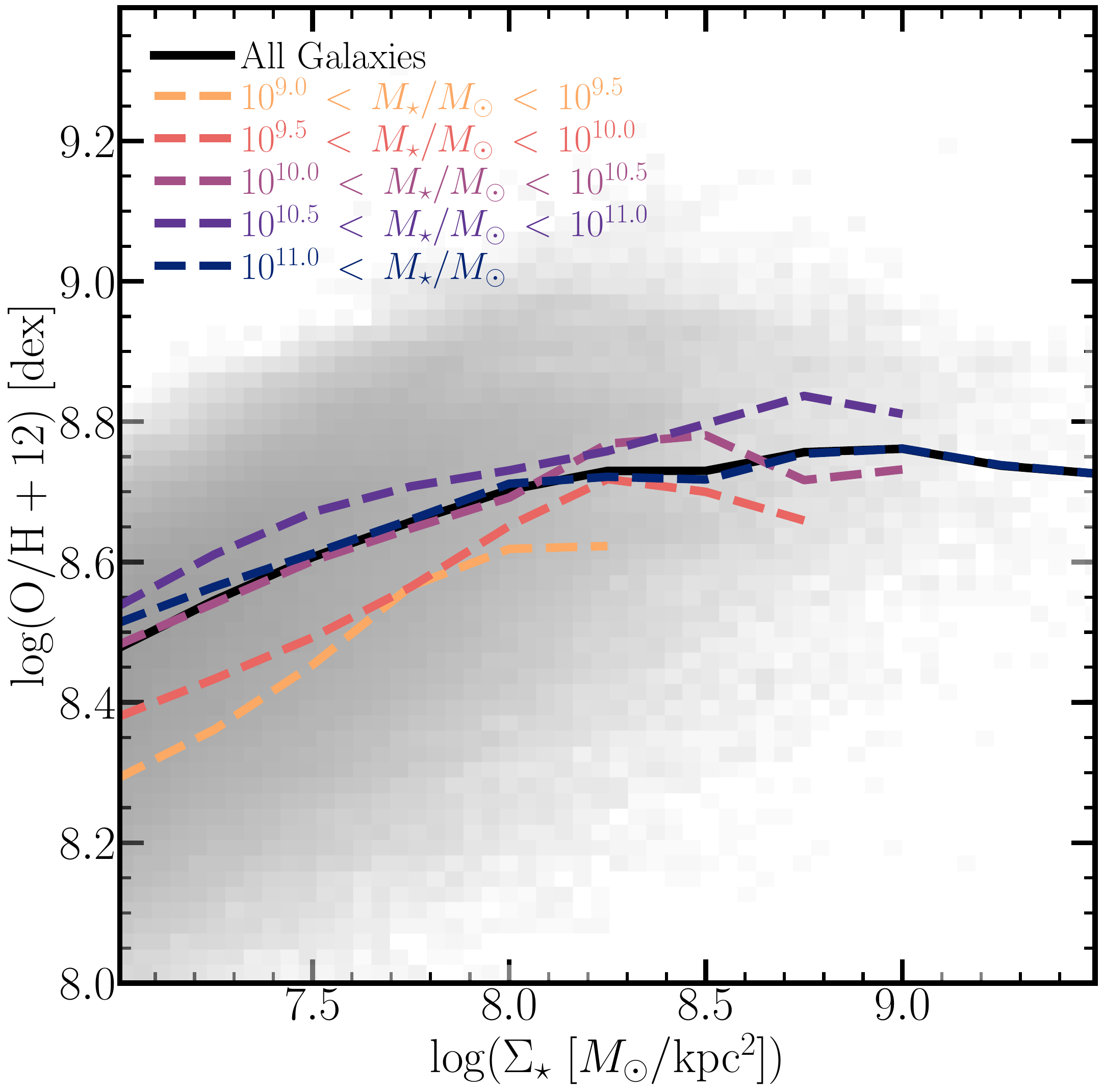}
        \else
            \includegraphics[width=\columnwidth]{bin_rmzr.pdf}
        \fi
    \else
        \includegraphics[width=\columnwidth]{bin_rmzr.pdf}
    \fi
    \caption{
    {\bf The rMZR for galaxies binned by total mass.}
    The rMZR split into different global stellar-mass bins (colored lines).
    The black solid line represents the median rMZR and the gray shaded region is the $1\sigma$ scatter (same as Figure~\ref{rzms}).
    The background histogram is the full population of spaxels.
    } 
    \label{rzms_bin}
\end{figure}

Figure \ref{rzms_bin} shows the rMZR, binned by host galaxies' total stellar mass.
The background histogram \edit{and solid} median line are the same as from Figure~\ref{rzms} while the \edit{dashed} colored lines represent the median rMZR for spaxels with \edit{different} host stellar masses.
\edit{There is a clear trend in the metallicity given the host mass of the galaxy: galaxies at a higher (total) stellar mass tend to have higher metallicities.
This trend of increasing metallicity with increasing stellar mass is a feature of the well-established mass-metallicity relation \citep[see][for broader discussion of the MZR in the TNG model]{Torrey_2018,Garcia_2024b,Garcia_2024c}.
The shape of the rMZR has only a weak dependence on the host stellar mass, with the lowest mass bins having a slightly steeper rMZR than the highest mass bins.
The difference in shape is likely related to the strength of metallicity gradients in the galaxies.
Metallicity gradients have been shown to be steeper in low mass galaxies than high mass galaxies in the TNG model \citep{Hemler_2021,Garcia_2023,Garcia_2025}.
Moreover, as we show explicitly in Section~\ref{subsec:origin_rSFMS} (Figure~\ref{rsfms_bin_rad}), the higher $\Sigma_\star$ pixels are closer to the center of the galaxy.
Therefore, the low mass systems have a steeper rMZR due to the outskirts (low $\Sigma_\star$) having significantly different metallicity values than the inner regions (high $\Sigma_\star$), compared to their high mass counterparts.
Finally, we note that the location of the turnover in stellar mass surface density ($\sim10^{8.5}M_\odot\,{\rm kpc}^{-2}$) is roughly the same in each of the host stellar mass bins (with the exception of the lowest mass bin, which has a turnover at $10^{8.0}M_\odot\,{\rm kpc}^{-2}$).
}

\subsubsection{(Lack of) Dependence on Star Formation Rate Surface Densities}

\begin{figure}
    \centering

    \ifRobinson
        \includegraphics[width=\linewidth]{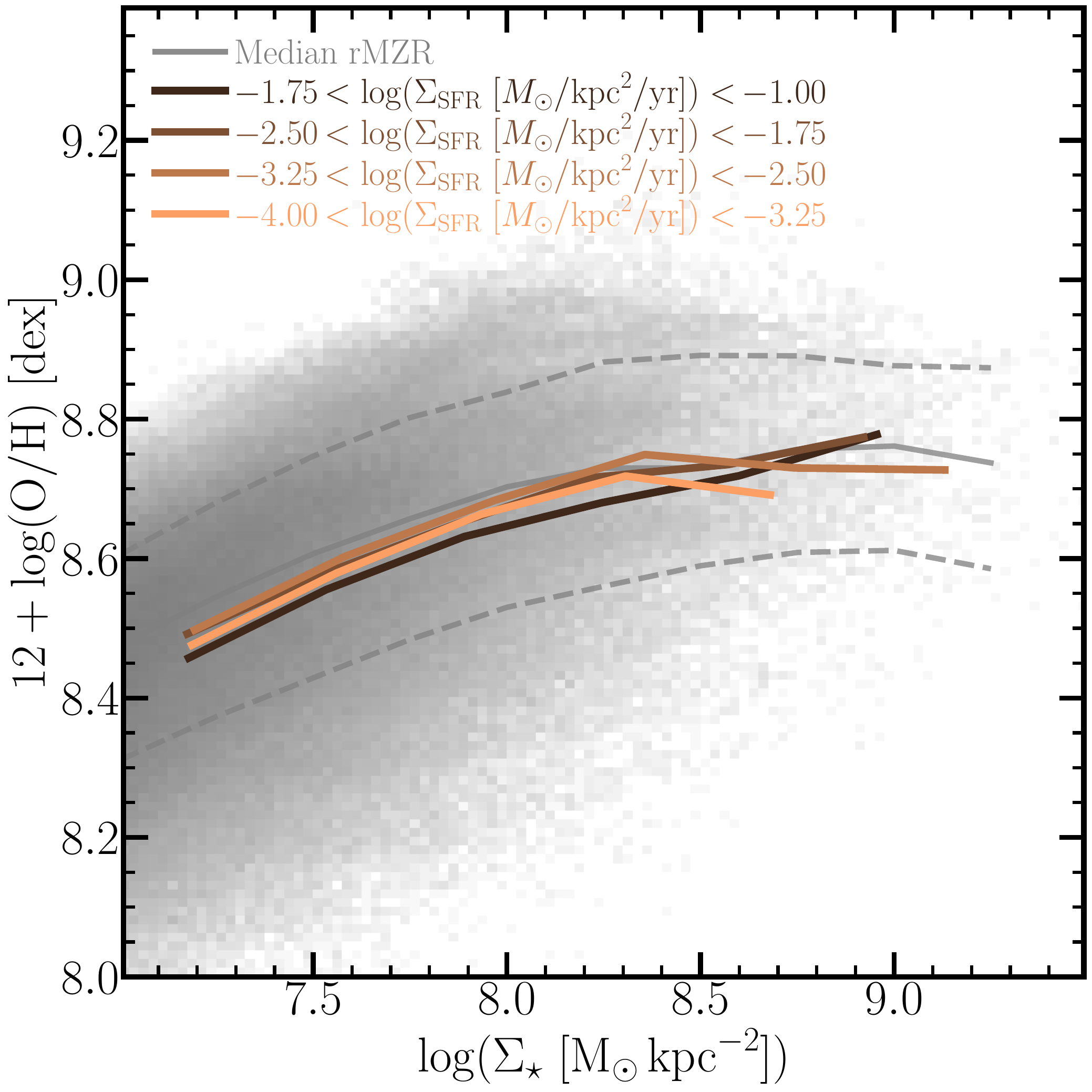}
    \else
        \includegraphics[width=\linewidth]{rFMR6.pdf}
    \fi
    \caption{{\bf The Resolved Fundamental Metallicity Relation (rFMR) in IllustrisTNG.}
    {\it Left:} The rMZR for different $\Sigma_{\rm SFR}$ bins. The 2D histogram in the background is the full distribution of metallicities from TNG (same as Figure~\ref{rzms}), while the gray lines are the median (solid) as well as 16$^{\rm th}$ and 84$^{\rm th}$ percentile (dashed) of the distribution.
    {\it Right:} The rMZR color-coded by the average spaxel's offset from the rSFMS. The inset shows the offsets from the rMZR against the offsets from the rSFMS, color coded by the average distance of the spaxels from the galactic center.
    }
    \label{fig:rFMR}
\end{figure}

In both observations \citep[e.g.,][]{Ellison_2008,Lara_Lopez_2010,Mannucci_2010,Belli_2013,Sanders2018} and simulations \citep{DeRossi_2015,Torrey_2018,Garcia_2024b,Garcia_2024c} the scatter about the global MZR is anti-correlated with the star formation rate.
This anti-correlation between a galaxy's SFR and metal content is oftentimes referred to as the Fundamental Metallicity Relation (or FMR).\ignorespaces
\footnote{\ignorespaces
~\!\!\edit{Despite the ``fundamental'' name, it is still debated, even at galactic scales, whether there exists a secondary correlation in the MZR with SFR \citep{Tremonti2004,Hughes_2013,Sanchez2013,Sanchez2017,Sanchez2019,Barrera2017,Garcia_2024b,Garcia_2024c}.
}}
Yet, it is unclear whether this ``fundamental'' relation holds on sub-galactic scales between $\Sigma_\star$, $\Sigma_{\rm SFR}$, and local metallicity (see \citeauthor{Sanchez2013} \citeyear{Sanchez2013}; \citeauthor{Barrera2016} \citeyear{Barrera2016}, \citeyear{Barrera2017}; \citeauthor{Sanchez-Menguiano_2019} \citeyear{Sanchez-Menguiano_2019}; \citeauthor{Trayford2019} \citeyear{Trayford2019}; \citeauthor{Ji_2022} \citeyear{Ji_2022}; \citeauthor{Schaefer_2022} \citeyear{Schaefer_2022}; \citeauthor{Baker_2023_FMR} \citeyear{Baker_2023_FMR}, \citeauthor{Koller_2024} \citeyear{Koller_2024}).

We present \edit{an analogous} {\it resolved} Fundamental Metallicity Relation (rFMR) from TNG in Figure~\ref{fig:rFMR}.
\edit{The background histogram shows the full distribution of pixels for the rMZR (i.e., that of Figure~\ref{rzms}).
The colored lines represent lines of constant $\Sigma_{\rm SFR}$ in bins of width $0.75$ dex.
}
We find that there is no \edit{strong} dependence on the SFR \edit{within the scatter of the rMZR}.
This result is potentially in tension with observations from \cite{Baker_2023_FMR}.
Those authors find that the rFMR follows the same trend as the integrated FMR in a MaNGA sample; though it should be noted that the anti-correlation with SFR appears to weaken with increasing $\Sigma_\star$ in that work.
Other observational works, however, seem to suggest there is no strong correlation between $\Sigma_{\rm SFR}$ and $Z_{\rm gas}$ at all (see, e.g., \citeauthor{Sanchez2013} \citeyear{Sanchez2013}; \citeauthor{Ji_2022} \citeyear{Ji_2022}).
It is therefore unclear the extent to which the \edit{lack of an} rFMR is in tension with observations.

\edit{
In all, we find that there is significantly more variation in the rMZR when considering the host stellar mass than variations in $\Sigma_{\rm SFR}$.
This suggests that the scatter about the rMZR in the TNG model is mostly driven by the properties of the host mass galaxy.
}
\section{Discussion}\label{sec:generalized_leaky_box}


\subsection{On the Origin of the rSFMS}
\label{subsec:origin_rSFMS}

\begin{figure}
    \centering
    \ifGarcia
        \ifRobinson
            \includegraphics[width=\linewidth]{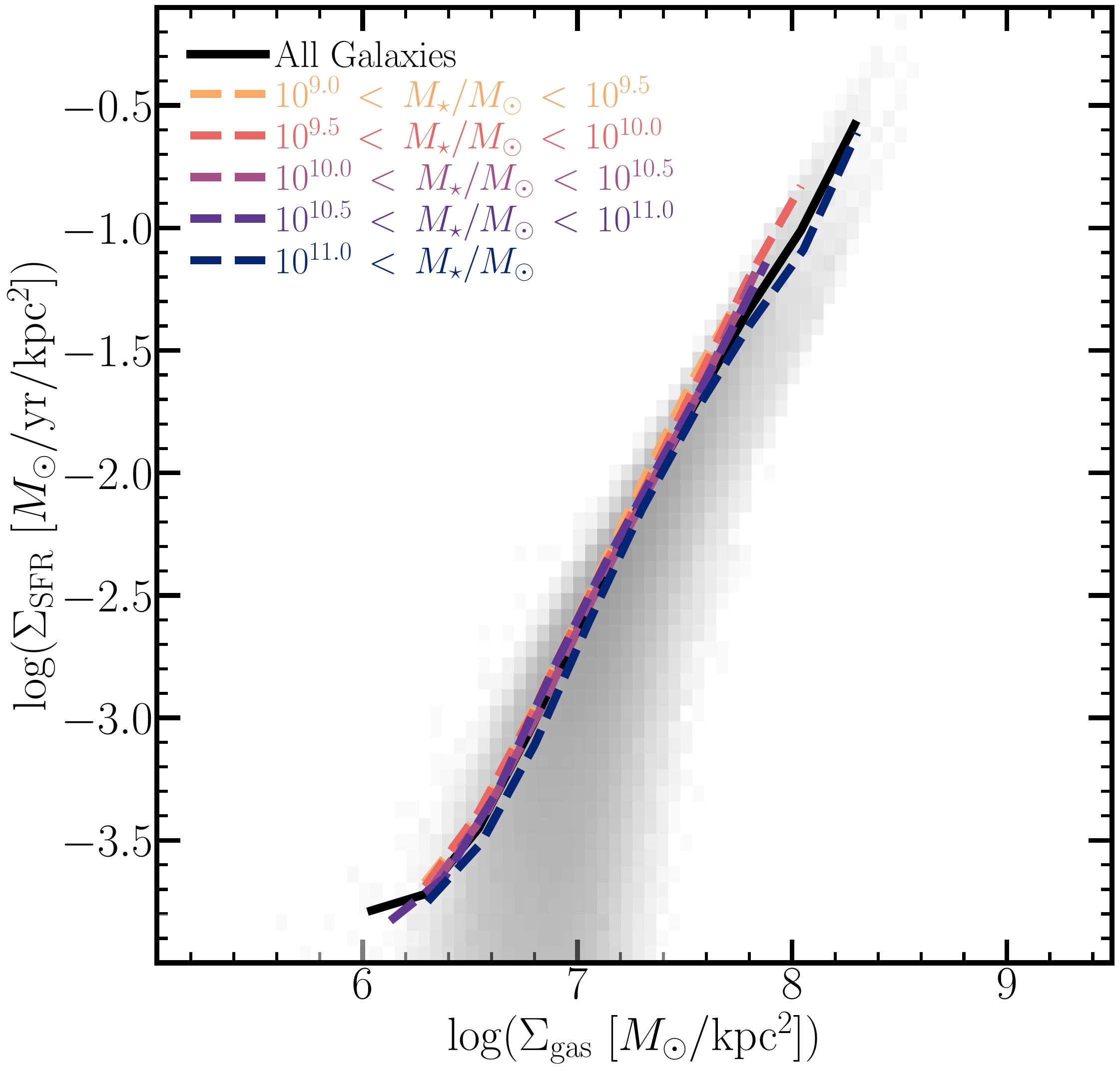}
        \else
            \includegraphics[width=\columnwidth]{ks_count.pdf}
        \fi
    \else
        \includegraphics[width=\textwidth]{ks_all.png}
    \fi
    \caption{{\bf Resolved Schmidt-Kennicutt (SK) relationship in TNG}. The SK relation for spaxels in TNG at $z=0$ with $1$ kpc spaxel resolution.
    The colored lines represent the median SK relation in different mass bins, while the black line represents the median relation including all spaxels.
    The background histogram is the full population of spaxels.
    }
    \label{ks}
\end{figure}

\begin{figure}
    \centering
    \ifRobinson
        \includegraphics[width=\linewidth]{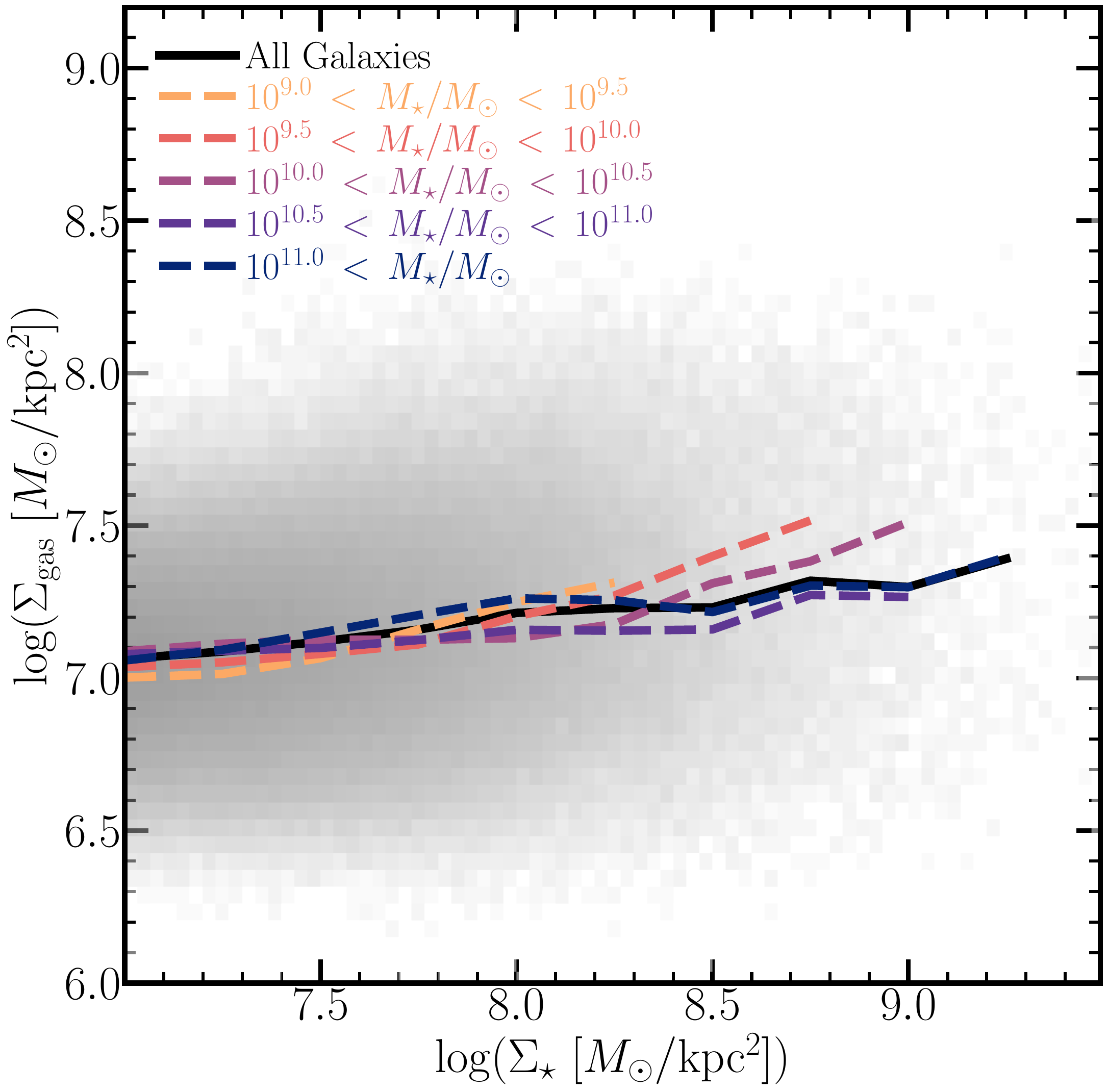}
    \else
        \includegraphics[width=\columnwidth]{ks_fgas_count.pdf}
    \fi
    \caption{{\bf The Gas Mass Main Sequence in IllustrisTNG} The $f_{\rm gas}-\Sigma_\star$ relation in TNG at $z=0$ with $1$ kpc spaxel resolution.
    The distribution in the background is the full population of spaxels from TNG and the solid black line is the median of this distribution.
    The dashed colored lines represent the median $f_{\rm gas}-\Sigma_\star$ relation in different stellar mass bins.
    }
    \label{fig:fgas_sigmastar}
\end{figure}

To understand the origin of the rSFMS, it is helpful to break down the rSFMS's dependence on both the star formation efficiency as well as gas content within galaxies.
The dependence on star formation efficiency roots in the Schmidt-Kennicutt (SK) relation, which is a well-established relation between $\Sigma_{\mathrm{SFR}}$ and $\Sigma_{\mathrm{gas}}$ with the form 
\begin{equation}
    \Sigma_{\mathrm{SFR}} = \epsilon \Sigma_{\mathrm{gas}}^{k}~,
    \label{eqs:ks1}
\end{equation}
where $\epsilon$ is star formation efficiency and $k$ is the power-law index \citep{Schmidt_1959,Kennicutt_1998}. 
Figure~\ref{ks} shows the median SK relation for all TNG spaxels (solid black line) as well as in the different mass bins (dashed colored lines).
In both the full population and within mass bins, there is a tight positive correlation between $\Sigma_{\mathrm{SFR}}$ and $\Sigma_{\mathrm{gas}}$ with \edit{$\epsilon = 5.89\times10^{-14}$ and $k = 1.52$, in agreement with previous results in TNG from \cite{Diemer_2018} and observations \cite[e.g.,][]{Kennicutt_1998}.
We emphasize that the rSK relation we present here is a {\it total} gas mass surface density, not molecular gas mass surface density.
The power-law index of $1.52$ would be too steep for a molecular gas mass SK \citep[see, e.g.,][their Table 1 for a summary of recent observations]{Sanchez_2023} and, indeed, \citeauthor{Diemer_2018} (\citeyear{Diemer_2018}; who build a model for different phases of the gas) find that the $\Sigma_{{\rm H}_2}-\Sigma_{\rm SFR}$ relation has a power-law index of $\sim1$ in the TNG model.
Regardless,} the tight correlation is expected behavior since the TNG model manually implements a version of the SK relation via the gas physical (3D) density (see Equation~\ref{eqn:volumetric_SK} and discussion in Section~\ref{sec:method_simulation}).
We therefore stress that the tightness \edit{and shape} of the ``resolved'' version of SK relation follows from this prescription \edit{and is not a prediction of the model}.
\edit{Within the individual mass bins, we find that the SK is universal across all mass bins.
This universality of the SK relation across mass bins in TNG is also a byproduct of the volumetric SK relation being implemented agnostic to host galaxy mass.
Star formation occurs in the same way across all mass bins in TNG.
This universal SK relation is potentially quite different to observed galaxies, which can have different depletion times at different stellar masses \citep[e.g.,][]{Colombo_2018} or different zero points for the SK relation at different stellar masses \citep[e.g.,][]{Sanchez_2021}.
}

Figure~\ref{fig:fgas_sigmastar} shows $\Sigma_{\mathrm{gas}}$ as a function of $\Sigma_{\star}$ (i.e., the ``\edit{total} gas mass main sequence'') for all TNG spaxels (in black) and different mass bins (in colors). 
We find a power-law trend between $\Sigma_{\mathrm{gas}}$ and $\Sigma_{\star}$ such that 
\begin{equation}
    \Sigma_{\rm gas} \propto \Sigma_\star^{n},
    \label{eqn:fgas}
\end{equation}
with \edit{$n=0.137$} in TNG.
\edit{Similar to the SK relation, there is very little dependence of the total gas mass main sequence on the host stellar mass of the galaxy.
The exception here is that high $\Sigma_{\star}$ pixels in the highest mass bins $(M_\star > 10^{10.5}M_\odot)$ have slightly depressed gas content.
}
We can combine Equations~\ref{eqs:ks1}~and~\ref{eqn:fgas} to obtain
\begin{equation}
    \Sigma_{\rm SFR} \propto \Sigma_{\star}^{nk}~.
    \label{eqs:ks3}
\end{equation}
Equation~\ref{eqs:ks3} is of the form of the rSFMS (i.e., Equation~\ref{eqn:rSFMS}) with $\alpha = nk$.
Combining the measured $n$ and $k$ from earlier in this section, we obtain a value of \edit{$\alpha = 0.20$}.
Admittedly, this approach yields a shallower power-law than measured in Section~\ref{sec:results_rsfms} (by around 33\%).
A source of this discrepancy may come from using a total gas mass main sequence opposed to a {\it molecular} gas mass main sequence.
\edit{Indeed, the difference in the power-law index for the {\it total} gas mass main sequence presented here ($n=0.137$) and that of {\it molecular} gas mass main sequence in observations ($\sim 1$; e.g., \citeauthor{Lin_2019} \citeyear{Lin_2019}) may be driven by difference in gas tracer.
It is possible that using the total gas mass in TNG is less fundamental as not all of the gas in the galaxy is star forming.}
Recent observational work has shown that the molecular gas content of galaxies is what is driving the existence of the rSFMS \citep{Lin_2019,Morselli_2020,Baker2022,Baker_2023_rSFMS,Ellison_2024}\ignorespaces
\footnote{\ignorespaces
Moreover, \cite{Ellison_2024} show that the dynamical equilibrium pressure of systems is even more fundamental in driving the rSFMS.
}.
The molecular gas content of galaxies is likely more tightly correlated with the ongoing SFR than the overall gas fraction since star formation occurs within giant molecular clouds.
The TNG model, however, does not explicitly track the individual phases (i.e., atomic, neutral, and molecular) of gas.
Efforts have been made to recover the different phases of the gas in post-processing (see, e.g., \citeauthor{Diemer_2019} \citeyear{Diemer_2018}, \citeyear{Diemer_2019}); regardless, we leave the detailed examinations of the individual phases of the gas for future work.

\subsubsection{On the Mass Dependence of the rSFMS}
\label{subsec:mass_rSFMS}

\begin{figure}
    \centering
    \ifRobinson
        \includegraphics[width=\linewidth]{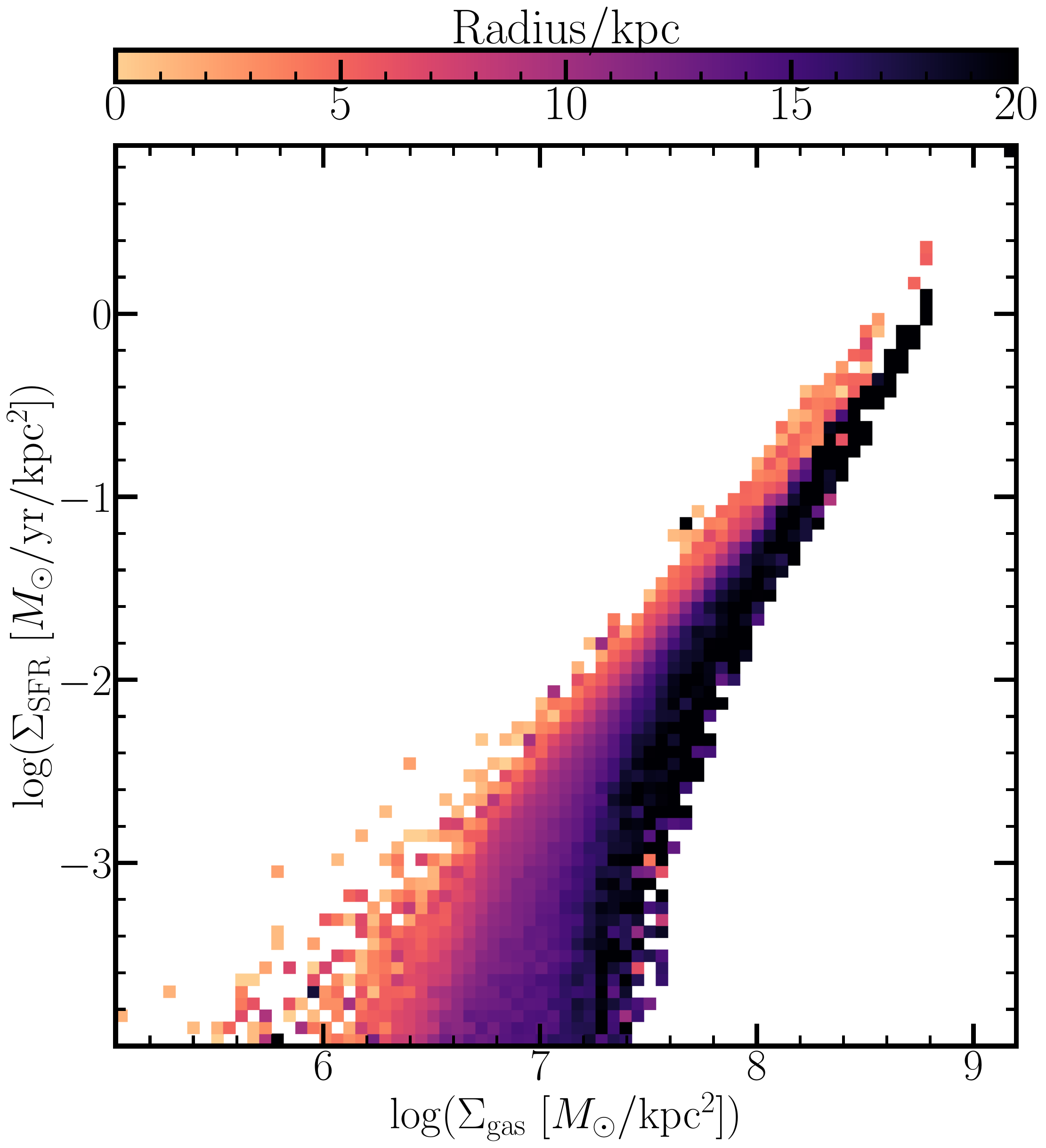}
    \else
        \includegraphics[width=\columnwidth]{ks_fgas_rad.pdf}
    \fi
    \caption{{\bf Radial dependence of the SK relation.} The SK (same as Figure~\ref{ks}) colored coded by the average galactocentric distance of spaxels.
    }
    \label{fig:ks_fgas_rad}
\end{figure}

\begin{figure*}
    \ifGarcia
        \centering
        \ifRobinson
            \includegraphics[width=\linewidth]{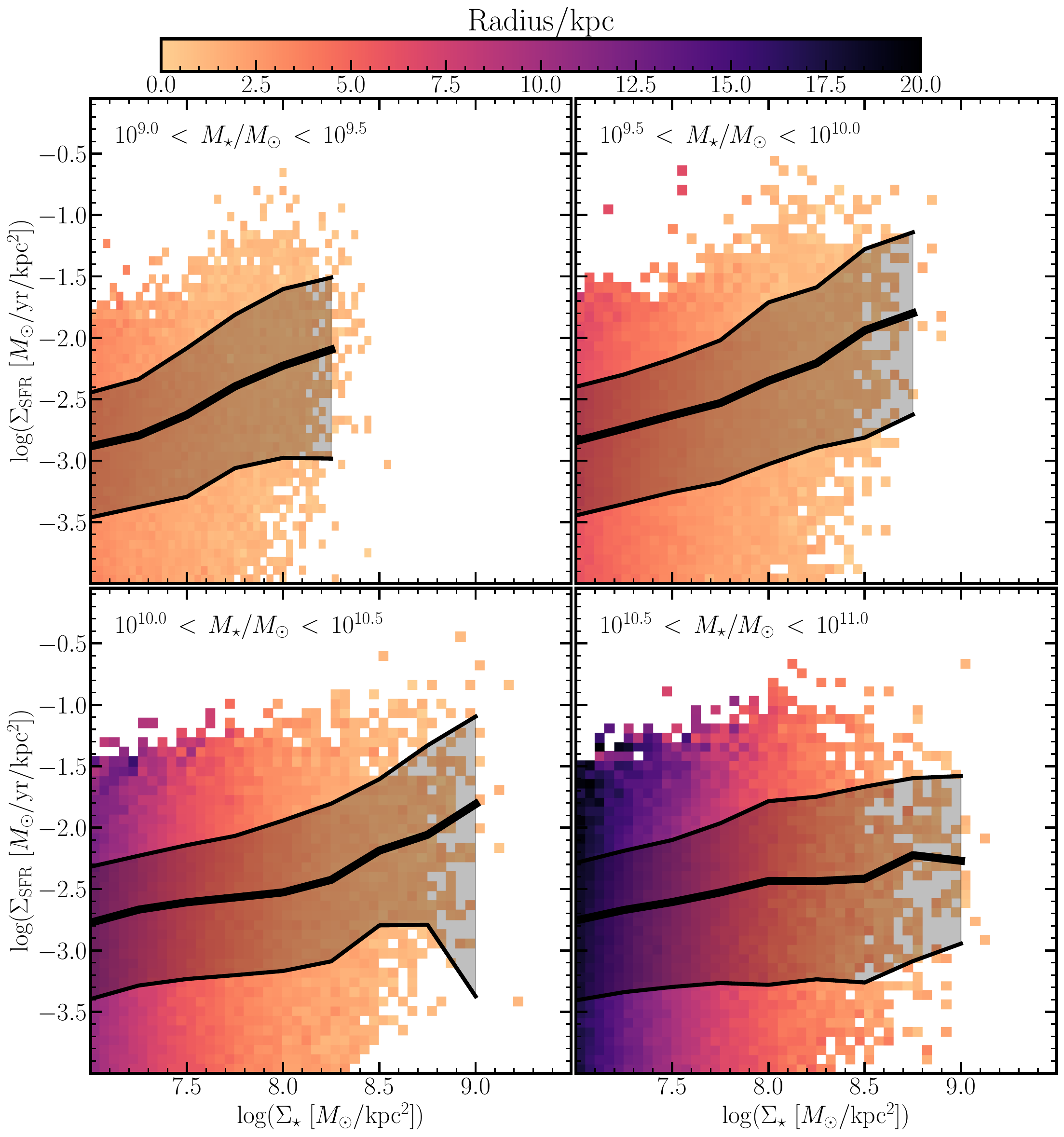}
        \else
            \includegraphics[width=\linewidth]{hist2d_sfr_rad.pdf}
        \fi
    \else
        \centering{
        \includegraphics[width=0.45\textwidth]{hist2d_sfr_mass090_rad.png}
        \includegraphics[width=0.45\textwidth]{hist2d_sfr_mass095_rad.png}}
        \centering{
        \includegraphics[width=0.45\textwidth]{hist2d_sfr_mass100_rad.png}
        \includegraphics[width=0.45\textwidth]{hist2d_sfr_mass105_rad.png}
        }
    \fi
    \caption{
    {\bf The radial dependence of the rSFMS.}
    Same as Figure~\ref{rsfms_com}, but now binned by total stellar mass of the host galaxy (top-left is $10^{9.0}<M_\star/M_\odot<10^{9.5}$, top-right is $10^{9.5} < M_\star/M_\odot < 10^{10.0}$, bottom-left is $10^{10.0} < M_\star/M_\odot < 10^{10.5}$, and bottom-right is $10^{10.5} < M_\odot < 10^{11.0}$).
    Additionally, each panel is color-coded by the average galactocentric radius of the spaxels.
    } 
    \label{rsfms_bin_rad} 
\end{figure*}


One counter-intuitive point to the above argument is that both the median SK relation and $\Sigma_{\mathrm{gas}}$-$\Sigma_{\star}$ relation show very little dependence on host galaxies' stellar masses. 
It should follow, then, that the rSFMS would not depend on host galaxies' stellar masses either, since the rSFMS is a combination of the two relations. 
Yet, there is some interesting, stellar mass dependent behavior hidden within the rSFMS (see Section~\ref{subsubsec:rSFMS_mass_TNG}, Figure~\ref{rsfms_com}).
In this section, we argue that the dependence of rSFMS on host galaxies' stellar masses comes from the combination of the SK relation and radially decreasing $\Sigma_\star$ profiles.

Figure~\ref{fig:ks_fgas_rad} shows the SK relation color-coded by the average galactocentric distance within each (histogram) pixel. 
There is a clear trend in the scatter of spaxels with distance to their galactic centers.
For spaxels with the same $\Sigma_{\mathrm{gas}}$, spaxels closer to their galactic center tend to have a higher $\Sigma_{\mathrm{SFR}}$.
Clearly, this shift in the normalization of SK relation with spaxels' radii indicates a change in the star formation efficiency for spaxels at different locations within the galaxies. 
For regions near the galactic center, gases are more compressed vertically compared with regions further away, since they sit deeper in the potential well of the dark matter halo.
Even with the same $\Sigma_{\mathrm{gas}}$, those central regions have a higher physical gas density $\rho_{\mathrm{gas}}$ and, consequently, a higher star formation efficiency. 
Moreover, the spaxels with large values of $\Sigma_{\star}$ tend to be closer to their galactic centers, which is unsurprising based on the radially-declining $\Sigma_{\star}$ profiles in TNG \citep{Wang2018,Wang_2023}.
We also note that the range of galactocentric distances depends on stellar mass: more massive galaxies are larger, and thus their spaxels span a wider range of distances from their respective centers \citep[see, e.g.,][]{Genel_2018}.

\edit{To put the radial dependence more concretely,} Figure~\ref{rsfms_bin_rad} shows the rSFMS broken into the four different mass bins and color-coded by the spaxel's galactocentric distance.
We find that spaxels which belong to less massive galaxies tend to be closer to the center of their host than spaxels of more massive galaxies at a fixed $\Sigma_\star$.
As discussed in the previous section, the regions which are closer to the center of the galaxy preferentially live in the upper right of the SK relation (i.e., high gas mass, high SFR; Figure~\ref{fig:ks_fgas_rad}).
Therefore, for a fixed $\Sigma_\star$, lower mass galaxies tend to have spaxels which are more star-forming than their high (total) mass counterparts.
This makes the rSFMS steeper for less massive galaxies than for more massive galaxies (seen both in Figures~\ref{rsfms_com}~and~\ref{rsfms_bin_rad} and Table~\ref{tab:rSFMS}). 
The dependence of rSFMS on host galaxies' stellar masses is therefore a consequence of changing star formation efficiency at different galactocentric distances for sub-galactic regions \edit{in the TNG model}.

\subsection{A Generalized Resolved Leaky-Box Model}

On global scales, closed box and/or leaky box (also referred to as bathtub) models have been employed to interpret the observed scaling relations \citep[see, e.g.,][]{Finlator2008,Lilly2013,Forbes2014}. 
Yet there has been comparatively little done on {resolved} scales.
One such work, by \cite{Zhu2017}, designs a {resolved} leaky-box model to describe the evolution of the $\Sigma_\star$, $\Sigma_{\rm gas}$, and gas-phase metallicity at the spaxel level.
The \cite{Zhu2017} model successfully predicts a tight rMZR, in line with the observed relation from IFS surveys \citep[see, e.g.,][]{Sanchez2013,Barrera2016}.
\edit{\cite{Barrera_Ballesteros_2018} extend the \cite{Zhu2017} model further by including the impact of gas fractions and escape velocities.
They find that while a gas-regulator framework reproduces the observed gas fraction-metallicity relation with physically motivated outflow rates, the leaky-box model alone requires unrealistically large mass loading factors to match the data.
}
Inspired by these results, we modify and expand the \edit{\cite{Zhu2017}} resolved leaky box model to include the $\Sigma_{\rm gas}-\Sigma_\star$ relation as well as explicit inflow and outflow terms. 

We take a rectangular prism through a $1~{\rm kpc}\times1~{\rm kpc}$ patch of the galaxy to represent an ``observed'' spaxel.
The evolution of this spaxel can be quantified in terms of its components: 
(i) stellar mass surface density $\Sigma_{\star}(t)$,
(ii) gas mass surface density $\Sigma_{\mathrm{gas}}(t)$,
(iii) SFR surface density $\Sigma_{\mathrm{SFR}}(t)$, and
(iv) local gas-phase metallicity $Z(t)$.

We begin with the evolution of stellar mass surface density.
To first order, the stellar mass content within our spaxel will be governed by the competition of new star formation and the death of old stars.
This can be expressed as
\begin{equation}
    \frac{{\rm d}\Sigma_\star}{{\rm d}t} = (1-{\cal R})\Sigma_{\rm SFR}(t)~,
    \label{eqn:Sigma_star}
\end{equation}
where ${\cal R}$ denotes the fraction of stellar mass that is returned to the ISM from short-lived massive stars.
In more detail, there should also be some term that describes the radial migration of stars into and out of the spaxel.
However, we assume that the average stellar mass migration is negligible \citep[see][]{Gurvich2020}.

We next consider the evolution of the gas content of our spaxel.
Similar to the stellar component, the gas is driven by consumption (new star formation) and enrichment (mass return from stars).
The gas mass is therefore, in a closed system, governed in a very similar way to Equation~\ref{eqn:Sigma_star} (albeit with an opposite sign):
\begin{equation}
    \frac{{\rm d}\Sigma_{\rm gas}}{{\rm d}t} = -(1-{\cal R})\Sigma_{\rm SFR}(t)~.
    \label{eqn:Sigma_gas}
\end{equation}
To extend Equation~\ref{eqn:Sigma_gas} to a leaky-box model, we must consider gas exchange with neighbouring spaxels.
In principle, gas can advect across the boundaries of the rectangular prism in a number of different ways.
The TNG model specifically implements gas outflows in the form of winds, whose strength is related to the SFR \citep{Springel2003}.
We can therefore write the surface density of outflowing gas as
\begin{equation}
    \frac{{\rm d}\Sigma_{\mathrm{gas,~out}}}{{\rm d}t} = -\eta_{\mathrm{out}} \Sigma_{\mathrm{SFR}}(t) ~,
    \label{eqs:box_outflow}
\end{equation}
where $\eta_{\mathrm{out}}$ is the outflow mass loading factor of these winds.
Similarly, previous studies suggest that the inflow rate is proportional to the SFR on galactic scales \citep{Toyouchi2015}.
We therefore make the (crude) assumption that this proportionality holds down to the sub-galactic scales such that the surface density of inflowing gas can be expressed as
\begin{equation}
    \frac{{\rm d}\Sigma_{\mathrm{gas,~in}}}{{\rm d}t} = \eta_{\mathrm{in}} \Sigma_{\mathrm{SFR}}(t)~,
    \label{eqs:box_inflow}
\end{equation}
where $\eta_{\mathrm{in}}$ is the inflow mass loading factor of the winds.
Combining Equations~\ref{eqn:Sigma_gas}-\ref{eqs:box_inflow}, the complete description of the gas evolution is given by
\begin{equation}
    \frac{{\rm d}\Sigma_{\rm gas}}{{\rm d}t} = -(1-{\cal R}-\eta_{\rm in}+\eta_{\rm out})\Sigma_{\rm SFR}(t)~.
    \label{eqn:sigma_gas_sfr}
\end{equation}
Recall, however, that $\Sigma_{\rm SFR}$ is a function of $\Sigma_{\rm gas}$ via the SK relation (Equation~\ref{eqs:ks1}) so that 
\begin{equation}
    \frac{{\rm d}\Sigma_{\rm gas}}{{\rm d}t} = -\epsilon(1-{\cal R}-\eta_{\rm in}+\eta_{\rm out})\Sigma_{\rm gas}^k(t)~.
    \label{eqs:box_gas_full}
\end{equation}
Although Eqn.~\ref{eqs:box_gas_full} is a non-linear differential equation, if $k > 1$ (as in the case for the SK relation), we can express its solution analytically:
\begin{equation}
    \Sigma_{\mathrm{gas}}(t) = \left(\Sigma_{\mathrm{gas,\, 0}}^{1-k} - (1-k)(1-{\cal R} - \eta_{\mathrm{in}} + \eta_{\mathrm{out}}) \epsilon t \right)^{\frac{1}{1-k}}~.
    \label{eqs:box_gas_final}
\end{equation}
Indeed, Equation~\ref{eqs:box_gas_final} is a generalization of Equation 6 of \cite{Zhu2017} with explicit inflow and outflow terms.

Finally, to model the local metallicity evolution, $Z(t)$, in each spaxel, we introduce $\Sigma_{\mathrm{metal}}(t)$ the surface density of gas-phase metals such that
\begin{equation}
    \Sigma_{\mathrm{metal}}(t) = Z(t) \Sigma_{\mathrm{gas}}~,
    \label{eqs:box_metal}
\end{equation}
which implies that 
\begin{equation}
    \frac{{\rm d}{Z}}{{\rm d}t} = 
    \frac{1}{\Sigma_{\rm gas}}\left[\frac{{\rm d}\Sigma_{\rm metal}}{{\rm d}t} - Z(t)\frac{{\rm d}\Sigma_{\rm gas}}{{\rm d}t}\right]~,
    \label{eqn:dZdt}
\end{equation}
via the quotient rule of derivatives.
There are four channels through which amount of gas-phase metals in a spaxel can be altered.
In the closed box, metals can either be consumed through star formation
\begin{equation}
    \frac{{\rm d}{\Sigma}_{\mathrm{metal,~ consume}}}{{\rm d}t} = -Z(t) \Sigma_{\mathrm{SFR}}(t)~,
    \label{eqs:metal_consume}
\end{equation}
or stars (which generate new metals) can enrich the ISM,
\begin{equation}
    \frac{{\rm d}{\Sigma}_{\mathrm{metal,~ enrich}}}{{\rm d}t} = y \Sigma_{\mathrm{SFR}}(t),
    \label{eqs:metal_enrich}
\end{equation}
where $y$ is the metal yield from the stars.
Extending this to a leaky-box, gas inflows can bring in metals from the circumgalactic medium,  intergalactic medium and/or, from neighboring spaxels, 
\begin{equation}
    \frac{{\rm d}{\Sigma}_{\mathrm{metal,~in}}}{{\rm d}t} = Z_{\mathrm{in}} \frac{{\rm d}\Sigma_{\mathrm{gas,~in}}}{{\rm d}t} = Z_{\rm in}\eta_{\rm in}\Sigma_{\rm SFR},
    \label{eqs:metal_inflow}
\end{equation}
where $Z_{\mathrm{in}}$ denotes metallicity of the inflowing gas.
Gas (with the same metallicity of the spaxel) can also be launched from the spaxel via winds
\begin{equation}
    \begin{split}
    \frac{{\rm d}{\Sigma}_{\mathrm{metal,\, out}}}{{\rm d}t} &= Z(t) \frac{{\rm d}\Sigma_{\mathrm{gas,~out}}}{{\rm d}t} + \Sigma_{\rm gas,~out}(t)\frac{{\rm d}Z}{{\rm d}t} \\
    &= -Z(t)\eta_{\mathrm{out}} \Sigma_{\mathrm{SFR}} + \Sigma_{\rm gas,~out}(t)\frac{{\rm d}Z}{{\rm d}t}~.
    \end{split}
    \label{eqs:metal_outflow}
\end{equation}
Combining Equations~\ref{eqs:metal_consume}-\ref{eqs:metal_outflow}, the evolution of $\Sigma_{\rm metal}$ is described by
\begin{equation}
    \begin{split}
    \frac{{\rm d}\Sigma_{\rm metal}}{{\rm d}t} = &[y - Z(t) + Z_{\rm in}\eta_{\rm in} - Z(t)\eta_{\rm out}]\Sigma_{\rm SFR}(t)~\\
    & +\Sigma_{\rm gas,~out}(t)\frac{{\rm d}Z}{{\rm d}t}~.
    \end{split}
    \label{eqn:Sigma_metal}
\end{equation}

Plugging in Equations~\ref{eqn:sigma_gas_sfr}~and~\ref{eqn:Sigma_metal} into Equation~\ref{eqn:dZdt} we obtain an analytic expression for the evolution of the local metallicity
\begin{equation}
    \frac{{\rm d}Z}{{\rm d}t} = \frac{\Sigma_{\rm SFR}(t)}{\Sigma_{\rm gas}(t) - \Sigma_{\rm gas,~out}(t)}\bigg(y-Z(t){\cal R} + \eta_{\rm in}[Z_{\rm in} - Z(t)]\bigg)~.
     \label{eqn:dZdt_full}
\end{equation}
Equation~\ref{eqn:dZdt_full} contains three ``terms'' representing the physics driving the metallicity evolution.
The first term is the coefficient outside the parentheses.
This coefficient represents a ``gas consumption timescale'': it is the rate at which gas is turning into stars divided by the total amount of gas left in the spaxel after an outflow event.
If the local SFR increases (or the spaxel undergoes a large outflow event), the rate of change of metallicity will also increase.
The terms in the parentheses of Equation~\ref{eqn:dZdt_full} represent contributions from the closed and leaky boxes.
The closed box term is $y - Z(t){\cal R}$.
In plain terms, if the stars return more metals into the ISM, the spaxel will increase its metallicity faster.
The second regime is that when the spaxel is already metal rich its metallicity will change more slowly.
Combined, the yield and return suggest that, in the limit of no exchange from one spaxel to another, the system will be entirely dominated by new star formation and prior enrichment.
The leaky-box term is $\eta_{\rm in}[Z_{\rm in} - Z(t)]$.
This represents a dilution term from the inflowing material: as the difference between the inflowing metallicity and current metallicity increases, so too does the overall rate of change of the metallicity (scaled by the mass loading factors).
In the limit of no new star formation or return from stars to the ISM, the change in the spaxel's metallicity will be dominated by inflowing material.
Notably, Equation~\ref{eqn:dZdt_full} does not depend on the outflows from the spaxel beyond decreasing the gas reservoir for future star formation.

Equations~\ref{eqn:Sigma_star},~\ref{eqs:box_gas_full},~and~\ref{eqn:dZdt_full} give the full set of differential equations that regulate the evolution of a spaxel.
Before proceeding, there are a few model parameters we adopt based on the TNG model.
TNG assumes no primordial stars and metals, so that $\Sigma_{\star}(0)=0$ and $Z(0) = 0$.
The initial gas mass surface density, however, we leave as a free parameter using values ranging from $10^{7.5}$ to $10^{10.0}~M_\odot~{\rm kpc}^{-2}$ in steps of 0.25 dex.
Additionally, we assume that the inflowing metallicity at any given time is pristine ($Z_{\rm in}=0$) from the intergalactic medium\ignorespaces
\footnote{
In reality, inflows can be pre-enriched recycled gas from the circumgalactic medium, metal rich flows from neighboring spaxels, or non-zero metallicity gas from anywhere in the environment.
The effect of a non-zero $Z_{\rm in}$ would be to decrease ${\rm d}Z/{\rm d}t$ (from the dilution term of equation~\ref{eqn:dZdt_full}).
Moreover, if $Z_{\rm in} > Z(t)$ (i.e., very metal rich inflows), the sign of the dilution term will change, but the overall behavior of the model will be largely the same as $Z_{\rm in} < Z(t)$.
}.
The return fraction, $\mathcal{R}$, can be estimated as $\mathcal{R}=0.5$ by integrating the \citeauthor{Chabrier2003} (\citeyear{Chabrier2003}; which TNG uses) initial mass function.
We take $k$ and $\epsilon$ of the SK relation from the best fit in Section~\ref{subsec:origin_rSFMS}.
Finally, the average metal yield from stars in TNG is $y=0.05$ \citep{Torrey2019}.

\subsubsection{Fiducial Closed Box Model}

\begin{figure}
    \centering
    \ifGarcia
        \ifRobinson
            \includegraphics[width=\linewidth]{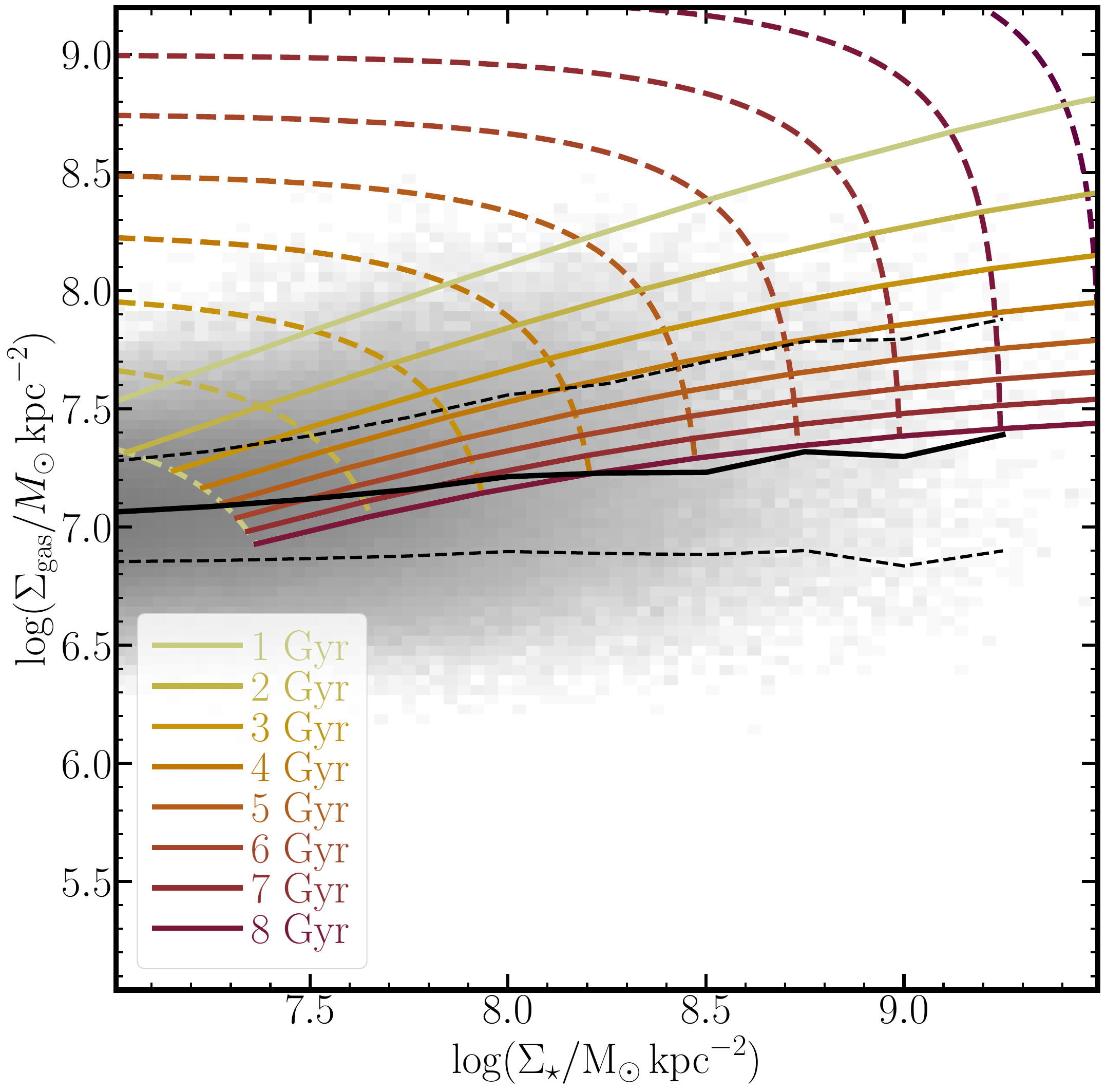}
        \else
            \includegraphics[width=\columnwidth]{closed_box_fgas.pdf}
        \fi
    \else
        \includegraphics[width=\columnwidth]{closed_box.png}
    \fi
    \caption{{\bf The gas mass main sequence in a resolved closed-box model.}
    evolution of simulated closed boxes on the MZR plane. The colored background display our IllustrisTNG spaxels. The dashed lines track the evolution of individual closed-boxed spaxels. The colored-solid lines our close-box predictions at various times, as labeled. 
    }
    \label{fig:box_time_series}
\end{figure}
\begin{figure}
    \centering
    \ifGarcia
        \ifRobinson
            \includegraphics[width=\linewidth]{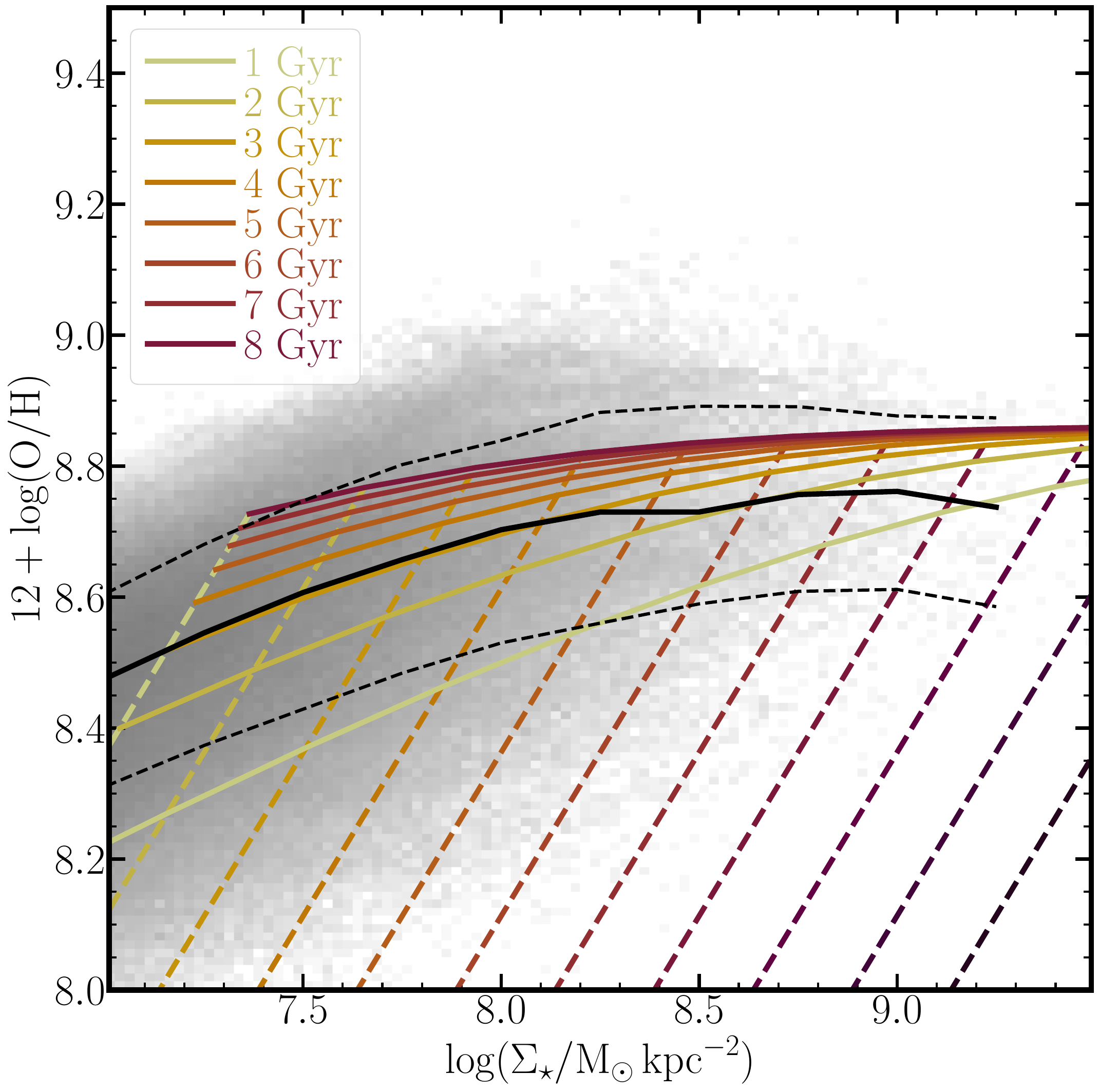}
        \else
            \includegraphics[width=\columnwidth]{closed_box_metal.pdf}
        \fi
    \else
        \includegraphics[width=\columnwidth]{closed_box_metal.png}
    \fi
    \caption{{\bf The rMZR in a resolved closed-box model.}
    The colored background shows the distribution of simulated TNG spaxels. The dashed lines track the evolution of representative individual spaxels. The solid lines represent the rMZR predictions in our closed-box model. }
    \label{fig:box_metal}
\end{figure}

We begin by examining our toy model using a closed box assumption where $\eta_{\rm in} = \eta_{\rm out} = 0.0$.
We numerically integrate Equations~\ref{eqn:Sigma_star},~\ref{eqs:box_gas_full},~and~\ref{eqn:dZdt_full} given the initial conditions set forth in the previous section.

Figure~\ref{fig:box_time_series} shows $\Sigma_{\mathrm{gas}}$-$\Sigma_{\star}$ relation against our resolved leaky-box model.
The solid lines connect the position of different leaky-box spaxels at a fixed integration time, where as the colored-dash lines track the gas-fraction evolution for a fixed initial $\Sigma_{\rm gas}$ values.
The shape of the dashed lines indicate the connection between gas consumption and stellar growth.
We note that, after $1$ Gyr of evolution (lightest yellow), our leaky-box prediction reaches the upper envelope of our 2D distribution drawn from TNG -- and after $\sim$3-4 Gyrs, the model lands on the local relation.
The median $\Sigma_{\rm gas}-\Sigma_{\star}$ relation (shown in the thick solid red line) of TNG starts to emerge as the evolution continues.
Moreover, past $t\gtrsim5~{\rm Gyr}$ the behavior of our model becomes asymptotic, modest changes in the length of the integration time have only modest changes in the gas fractions.

Figure~\ref{fig:box_metal} shows our closed box model predictions for the rMZR.
The solid lines again indicate spaxel locations in the rMZR at times $t>t_0$ and the dashed lines depict the same five representative spaxel tracks displayed in Figure~\ref{fig:box_time_series}. 
As time progresses, a given spaxel grows its stellar component -- and it becomes more metal-rich, as gas recently locked in short-lived stars is returned to the ISM.
After $1$ Gyr of evolution, our prediction lies on the lower envelope of the TNG relation, reaching the local relation after $\sim$2-3 Gyr, noting that the closed box model assumptions made here seem to overpredict the median rMZR past $\sim3$ Gyr.
Similar to the behavior of the $\Sigma_{\rm gas}-\Sigma_{\star}$ relation, past $t\gtrsim$ 5 Gyr the behavior of this model is asymptotic.

\subsubsection{Inflow/Outflow Variations}

\begin{figure}
    \centering
    \ifGarcia
        \ifRobinson
            \includegraphics[width=\linewidth]{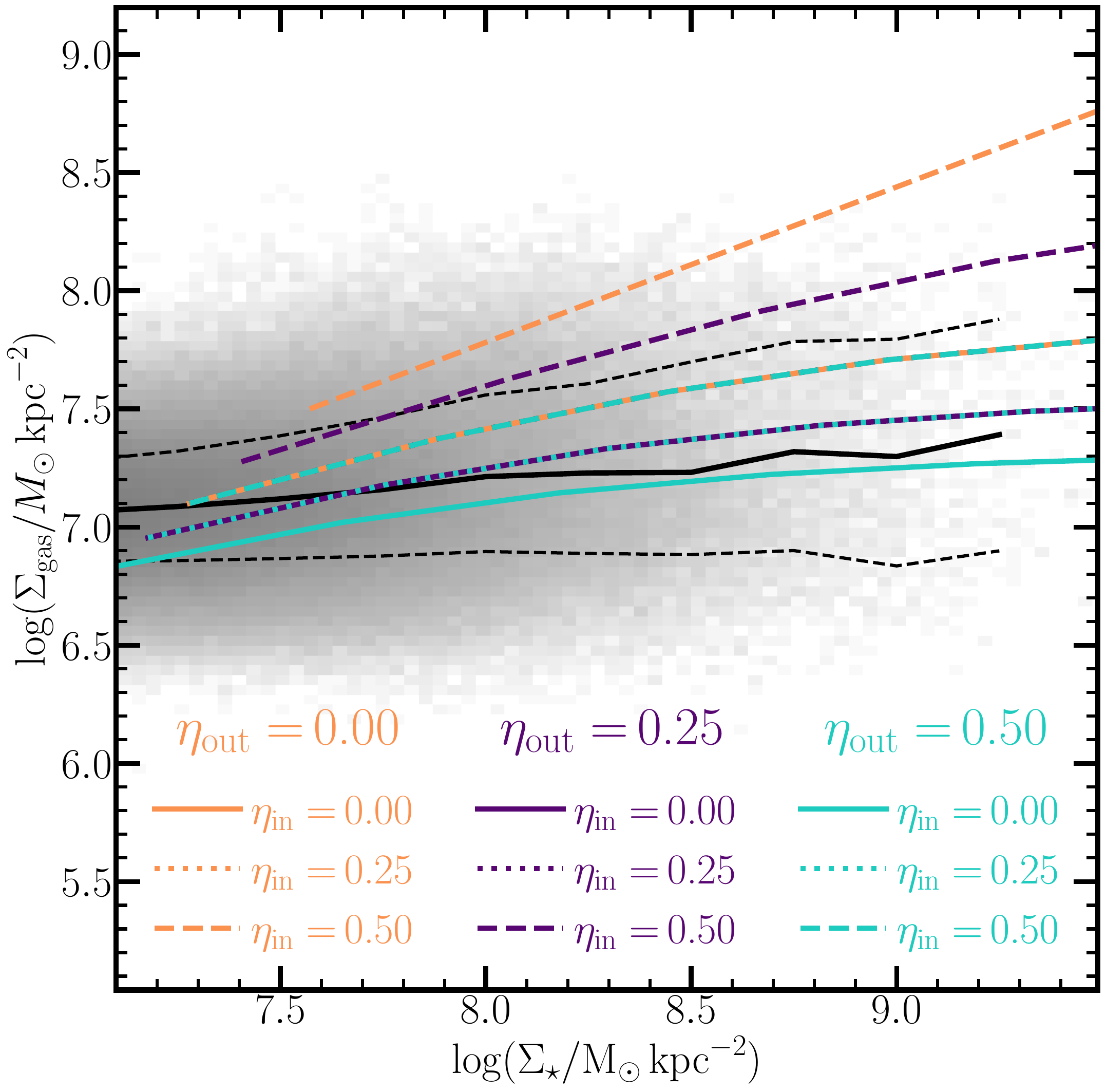}
        \else
            \includegraphics[width=\columnwidth]{closed_box_eta.pdf}
        \fi
    \else
        \includegraphics[width=\columnwidth]{closed_box_eta.png}
    \fi
    \caption{{\bf The gas mass main sequence in a resolved leaky-box model.}
    The $f_{\mathrm{gas}}$-$\Sigma_{\star}$ formed with different choices of $\eta_{\mathrm{in}}$ and $\eta_{\mathrm{out}}$. The colored background shows the distribution of actual IllustrisTNG spaxels included in this study. The colored-solid lines are the relations formed by spaxels at $t=5$ Gyr under different model assumptions, as labeled.
    }
    \label{fig:various_eta}
\end{figure}
\begin{figure}
    \centering  
    \ifGarcia
        \ifRobinson
            \includegraphics[width=\linewidth]{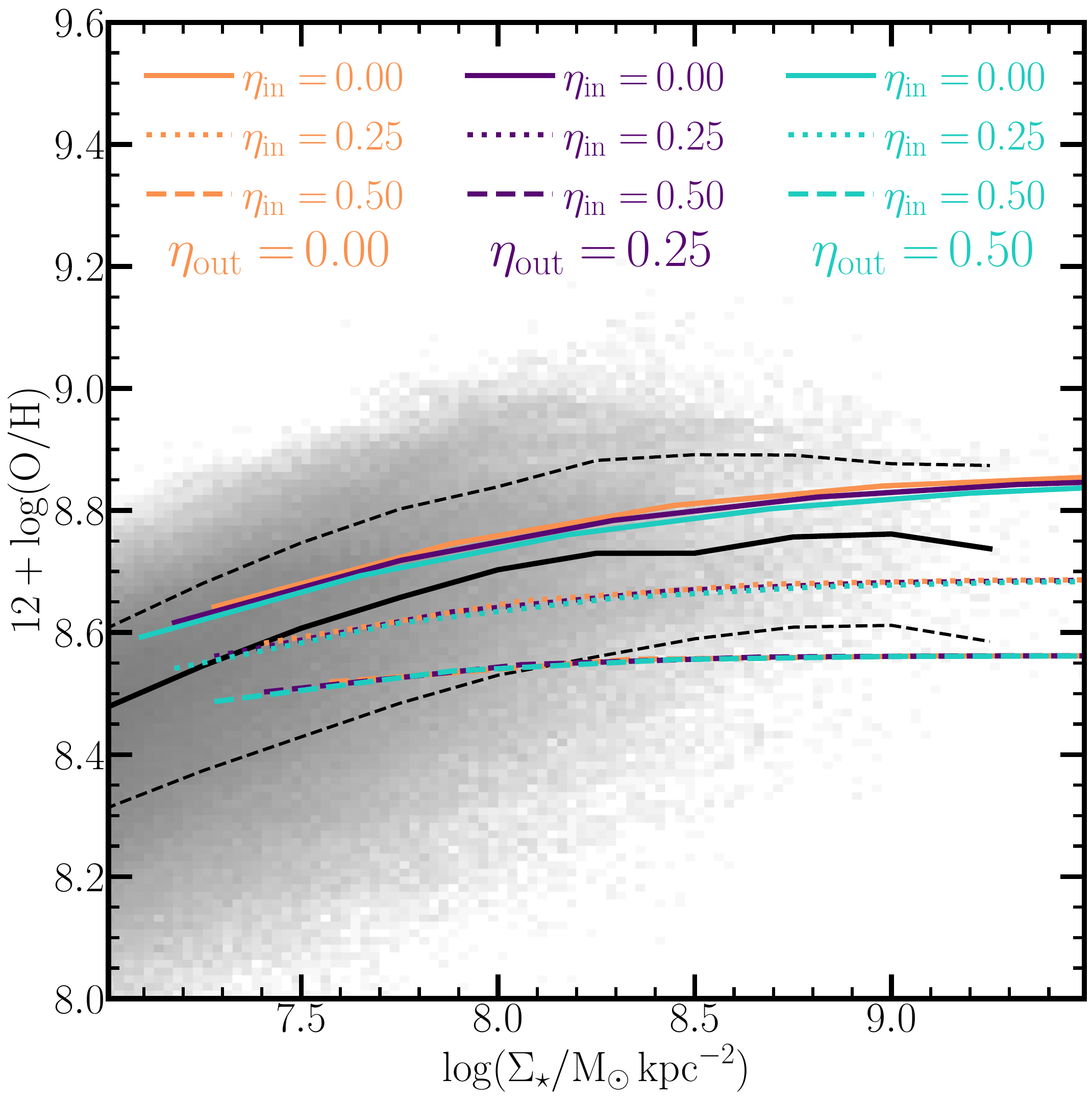}
        \else
            \includegraphics[width=\columnwidth]{closed_box_metal_eta.pdf}
        \fi
    \else
        \includegraphics[width=\columnwidth]{closed_box_metal_eta.png}
    \fi
    \caption{{\bf The rMZR in a resolved leaky-box model.}
    The rMZR formed with different choices of $\eta_{\mathrm{in}}$ and $\eta_{\mathrm{out}}$. The colored background shows the distribution of actual IllustrisTNG spaxels included in this study, the same as Figure~\ref{rzms}. The colored-solid lines are the relations formed by spaxels at $t=5$ Gyr under different model assumptions, as labeled.}
    \label{fig:various_eta_z}
\end{figure}

To explore how different inflow/outflow models affect the results for resolved scaling relations, we run the same leaky-box model with different mass-loading factors for inflows and outflows, namely, a combination of $\eta_{\mathrm{in}}=\{0.0, 0.25, 0.5\}$ and $\eta_{\mathrm{out}}=\{0.0, 0.25, 0.5\}$. 
The choice of upper bound value 0.5 for our mass-loading factor variation is because we adopt a stellar return fraction $\mathcal{R} =0.5$ and any inflow with $\eta_{\mathrm{in}} > 0.5$ will become a runaway process. 
Figure~\ref{fig:various_eta} and \ref{fig:various_eta_z} show predictions for rSFMS and rMZR from our resolved leaky-box model at $t=5$ Gyrs, after which the evolution of resolved scaling relations in our model becomes asymptotic, with different combinations of $\eta_{\mathrm{in}}$ and $\eta_{\mathrm{out}}$ values. 
The leaky-box $\Sigma_{\rm gas}-\Sigma_{\star}$ prediction relies on the net outflow. 
If the net outflow is negative, in other words, a net inflow, the spaxels' gas fraction becomes too high compared to what we observe for TNG spaxels. 
The supply of inflow gas maintains the gas fraction, and consequently, the SFR at high levels. 
To be consistent with the result for $\Sigma_{\rm gas}-\Sigma_{\star}$ relation we observe in TNG, a positive, or near zero, net outflow is preferred. 

On the other hand, the rMZR does not display dependence on the net inflow/outflow rate. 
Instead, models are affected by only the inflow factor $\eta_{\mathrm{in}}$: the higher the $\eta_{\mathrm{in}}$ is, the lower normalization and shallower slope the predicted rMZR has. 
This is because metallicity will be significantly diluted with the inflow of primordial gas, which has zero metallicity.
At fixed $\eta_{\mathrm{in}}$, the predicted rMZR is insensitive to the adopted outflow strength $\eta_{\mathrm{out}}$. 
This, obviously, stems from our assumption that the outflowing metallicity is the same as the current metallicity; therefore, the overall ratio of metal mass to total mass remains unchanged from outflows (see Equation~\ref{eqn:dZdt_full}).
If the metallicity of the outflows were preferentially metal poor or metal rich, however, the there would likely be an impact of the overall metal evolution of spaxels.
Regardless, resolved leaky-box models \citep[e.g.,][]{Zhu2017} that invoke `net' outflow rates (without a distinct inflow rate) do not sufficiently capture the behavior of the rMZR. 
\edit{The conclusion that inflows and outflow matter is qualitatively similar to the findings of \cite{Barrera_Ballesteros_2018}.
While those authors found that a leaky box model requires unphysically large outflows; their gas regulator model (which explicitly incorporates inflows/outflows) performs much better.
}

\edit{Overall, our results suggest that a generalized leaky-box framework can reasonably explain the simulated spaxels and highlights the importance of explicit inflow and outflow terms when considered resolved scaling relations.
However, we recognize that observational constraints \citep{Barrera_Ballesteros_2018} favor gas-regulator models with feedback tied to physical parameters such as escape velocity.
Future constraints from the observational perspective like those of \cite{Barrera_Ballesteros_2018} can provide strong bounds on the implementation of feedback in future simulation models.
}

\subsection{Limitations and Extensions of this Work}

\edit{
In all of the results of this paper, we quote the raw output from the simulations.
Observations, on the other hand, necessarily include extra steps of interpreting the emission lines from the galaxy spectra \citep{Brinchman2004,Kewley2008} and are limited by spatial resolution, signal-to-noise, and selection effects inherent in survey design and instrumentation \citep[e.g.,][]{Sanchez2012, Bundy2015}.
The observational limitations can significantly affect derived quantities such as star formation rates, metallicities, and kinematics, making direct comparisons between simulations and observations non-trivial.
The main advantage of using the raw simulation outputs in this way is to avoid any of these observational effects; however, omits several of the important features of observational data making strong comparisons between the observed and simulated datasets more difficult.
Therefore, a limitation of this work is that we do not forward model the galaxy observations.
}

\edit{
There has been much work recently to forward model the IFU data cubes from galaxy simulations (see, e.g,. {\sc simspin} \citeauthor{Harborne_2020} \citeyear{Harborne_2020}; {\sc galcraft} \citeauthor{Wang_2024} \citeyear{Wang_2024}; {\sc rubix} \citeauthor{Cakir_2024} \citeyear{Cakir_2024}; {\sc MANGiA} \citeauthor{Sarmiento_2023} \citeyear{Sarmiento_2023}; and \citeauthor{Ibarra-Medel_2019} \citeyear{Ibarra-Medel_2019}).
These tools simulate the full process of observation by generating mock IFU data cubes that incorporate radiative transfer, instrument effects, and survey-specific limitations such as spatial resolution, seeing, and noise characteristics. 
Forward modeling in this way allows for more direct comparisons between simulations and observational datasets, ensuring that the derived properties are interpreted within the same observational framework. 
Incorporating such methods in future work could be an important step toward robustly linking theoretical predictions with IFU survey data and for disentangling physical signals from observational biases.
}

\section{Summary}\label{sec:summary}

We have studied the spatially resolved scaling relations between physical properties for star-forming galaxies at $z\,=\,0$ in IllustrisTNG.
We selected star-forming galaxies from an IllustrisTNG snapshot and produce spatially resolved property maps for them (see Figure~\ref{fig:doodle}).
We then present the two scaling relations, the resolved star-formation main sequence (rSFMS) and the resolved mass-metallicity relation (rMZR), compare them with results from observational surveys, and discuss the tightness and residual of the rSFMS.
Our conclusions are as follows, 
    
\begin{enumerate}[label=(\roman*),leftmargin=0.5cm]
    \item The rSFMS exists within IllustrisTNG galaxies, as a positive single-power-law correlation between the stellar mass surface density $\Sigma_{\star}$ and the star formation rate surface density $\Sigma_{\mathrm{SFR}}$ at the scale of $1\,\mathrm{kpc}^{2}$ (Figure~\ref{rsfms}). 
    The rSFMS has a slope of $0.3$, and a $1\sigma$ scatter of $0.25$ dex. 
    The rSFMS from our results agrees well with IFU observations in both \edit{scatter} and normalization, \edit{but disagree significantly on the power-law index}.
    We find that the \edit{scatter about the} rSFMS \edit{in TNG is mainly driven by the scatter due to individual galaxies and does not strongly depend on host halo mass} (Figures~\ref{fit_all}~and~\ref{rsfms_com}).

    \item \edit{We find that the TNG model predicts that} gas-phase metallicity is positively correlated with $\Sigma_{\star}$ (Figure~\ref{rzms}). 
    The rMZR within IllutrisTNG galaxies qualitatively agrees with the similar metallicity relation from IFU observations (though we caution against too strong a comparison; see discussion in Section~\ref{sec:results_rmzr}). 
    In particular, the rMZR in IllustrisTNG successfully recreates the characteristic features of a single-power-law in the low mass regime and a plateau in the high mass regime. 
    \edit{We also do not find a significant resolved fundamental metallicity relation in the scatter of the rMZR (Figure~\ref{fig:rFMR}). 
    Rather, most of the scatter in the rMZR seems more correlated with host stellar mass (Figure~\ref{rzms_bin}).}
    

    \item We investigate the origin of the rSFMS by considering the Schmidt-Kennicutt relation and gas mass main sequence ($\Sigma_{\rm gas}$-$\Sigma_\star$ relation) \edit{in TNG} (Figures~\ref{ks}~and~\ref{fig:fgas_sigmastar}, respectively). 
    The combination of these two scaling relations can naturally explain the emergence of the rSFMS in TNG.
    
    \item Finally, we show that gas mass main sequence and rMZR for IllustrisTNG regions can be well described by a generalization of the leaky-box model of \citeauthor{Zhu2017} (\citeyear{Zhu2017}; see Figures~\ref{fig:box_time_series}-\ref{fig:various_eta_z}). 
    The gas mass main sequence depends on both the inflow and outflow mass loading factors, whereas the rMZR only exhibits a strong dependence on the inflowing mass loading factor.
    Overall the leaky-box model indicates that the resolved galaxy scaling relations in IllustrisTNG are governed by feedback on the local scales ($\sim1$ kpc) of the galaxy.
\end{enumerate}

Fundamentally, star formation and chemical enrichment are ``local'' processes within galaxies.
Understanding and quantifying the extent to which simulation models -- which nominally reproduce realistic galaxy populations on integrated scales -- agree/disagree with observations on these scales can be a critical test of their sub-grid prescriptions.
Current and future IFS campaigns promise to shed light in this area and give us a deeper understanding of what drives galaxy's evolution on sub-galactic scales.

\section*{Data Availability}

\edit{
Raw data products from the TNG simulations is publicly available. IllustrisTNG: \href{https://www.tng-project.org/data/}{https://www.tng-project.org/data/}.
We provide the scripts to generate the reduced data products and the results that support the conclusions in this analysis publicly via Github: \href{https://github.com/AlexGarcia623/TNG_Resolved_Scaling_Relations/tree/main}{github.com/AlexGarcia623/TNG\_Resolved\_Scaling\_Relations}
}

\begin{acknowledgments}
\edit{The authors acknowledge Research Computing at The University of Virginia and University of Florida Information Technology for providing computational resources and technical support that have contributed to the results reported within this publication.}

\edit{AMG and PT acknowledge support from NSF-AST 2346977 and the NSF-Simons AI Institute for Cosmic Origins which is supported by the National Science Foundation under Cooperative Agreement 2421782 and the Simons Foundation award MPS-AI-00010515.}
\end{acknowledgments}

\appendix

\section{Modified Smoothing Kernel}
\label{appendix:density_reconstruction}

\begin{figure}
    \centering
    \includegraphics[width=\linewidth]{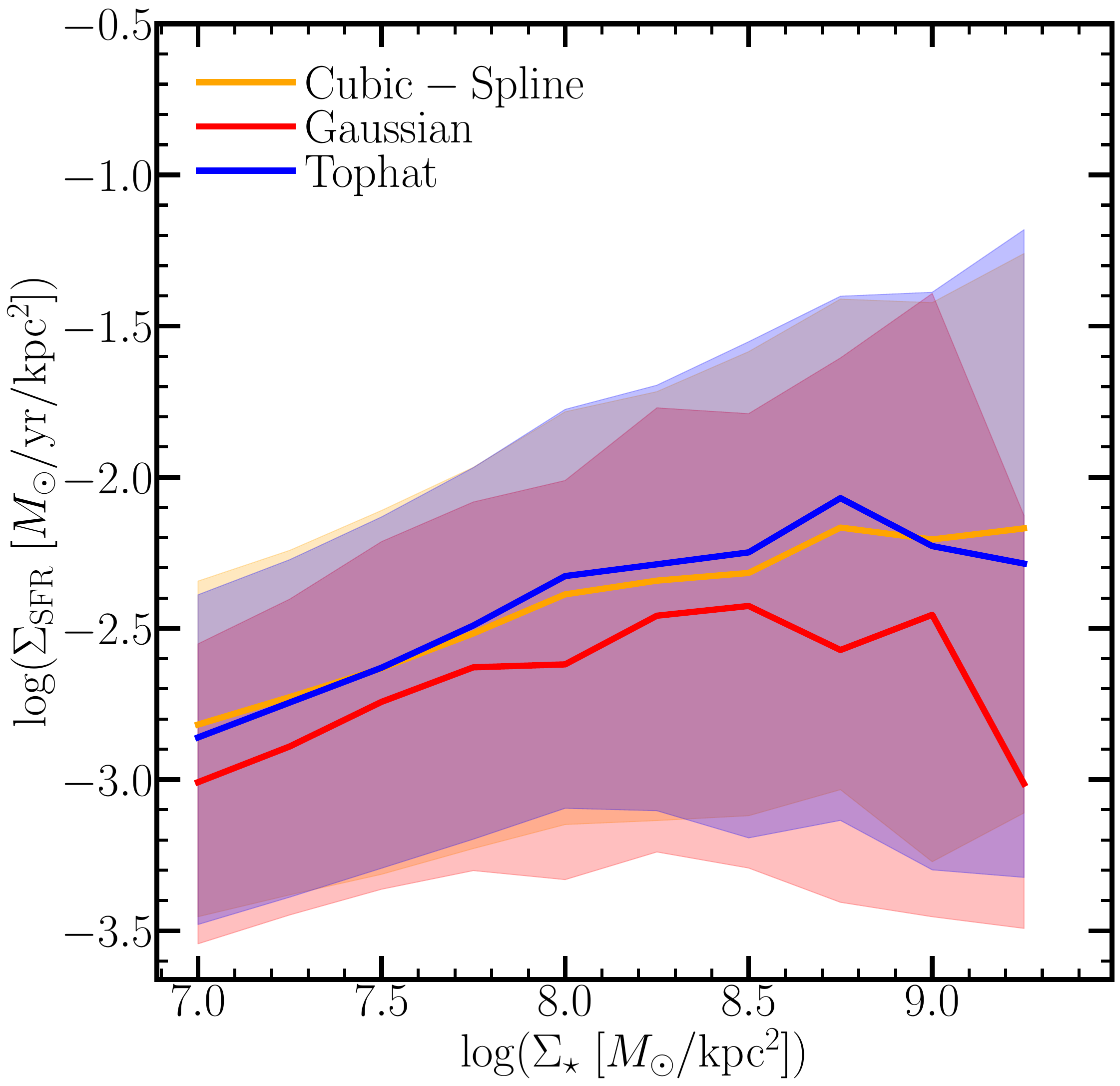}
        \caption{{\bf Modified Smoothing Kernel}}
    \label{fig:kernel_function}
\end{figure}

\edit{
In Section~\ref{sec:method_map}, we discuss our method for constructing resolved maps of the TNG galaxies. 
Throughout the main text, we utilize the cubic-spline kernel smoothing function to ``reconstruct'' the density by distributing the mass of each particle. 
The decision to use the cubic-spline kernel is heavily influenced by its widespread use in smoothed-particle hydrodynamics (SPH; \citeauthor{Gingold_Monaghan_1977} \citeyear{Gingold_Monaghan_1977}, \citeauthor{Monaghan_1992} \citeyear{Monaghan_1992}, \citeauthor{Springel_2010b} \citeyear{Springel_2010b}).
However, the choice of kernel weighting function is, in principle, arbitrary. 
To evaluate the sensitivity (or lack thereof) of our results to this choice, we also consider two alternatives; the gaussian and tophat kernels.
}

\edit{
In the SPH kernel weighting approach, the density at a given point is computed such that
\begin{equation}
        \rho_i = \sum_{j}^{N_{\mathrm{ngb}}} m_j \, W_{i,j}(r_j \,|\, h) ,
\end{equation}
where $m_j$ is the mass of the $j^{\rm th}$ neighboring particle, $W_{i,j}$ is the kernel weighting function, $r_j$ is the distance between the $i^{\rm th}$ point to the $j^{\rm th}$ particle, and $h$ is the smoothing length of the kernel.
In our case, the kernel function $W_{i,j}$ is averaged over three dimensions and has units of inverse volume.

The cubic-spline kernel is utilized throughout the main text, which takes this form 
\begin{equation}
    W = \frac{8}{\pi h^3}
    \begin{cases}
        1-6x^2 + 6x^3 & x \leq 0.5 \\
        2(1-x)^3 & 0.5 < x \leq 1.0 \;,\\
        0 & {\rm otherwise} 
    \end{cases}
\end{equation}
where the smoothing length, $h$, is constructed by searching the nearest 32 neighbors, and $x$ is the distance $r$ normalized by the smoothing length.
The alternatives (top-hat and gaussian) can be modeled using 
\begin{equation}
    W = \begin{cases}
        \frac{3}{4\pi h^3} & |x| \leq 1 \\
        0 & {\rm otherwise}\end{cases} \quad ,
\end{equation}
and
\begin{equation}
    W = \frac{1}{(\pi h^2 )^{3/2}} {\rm exp}
    \left(-\frac{|x|^2}{h^2}\right) \quad ,
\end{equation}
respectively. 
}
    
\edit{
Figure \ref{fig:kernel_function} shows the rSFMS obtained using each of these three methods for the same sample of TNG galaxies at $z = 0$. 
For each curve, the median is given by the solid line, and the scatter is represented by the shaded region (the upper and lower bounds representing the 84th and 16th percentile of the $\log\Sigma _{\rm SFR}$ at each $\log\Sigma_ \star$ bin, respectively).
We find that the choice of kernel has only a minor impact on the resulting rSFMS. 
The median relations for the three kernels agree within $\leq 0.25$ dex across the full range of $\Sigma_\star$, with small deviations appearing only at the highest surface densities.
}

\edit{
The gaussian kernel yields a slightly lower overall normalization.
The lowered normalization is due in large part to the implementation of the kernel function, which truncates at three standard deviations for numerical stability.,
Therefore some small ($\lesssim1\%$) amount of the kernel is lost from the distribution tails.
While a shimmeringly small effect for an individual particle's kernel shape, it can bias our estimate of the local density to preferentially lower values on aggregate.
By contrast, the tophat kernel tends to overestimate the median at higher $\Sigma_ \star$ compared to the cubic-spline kernel.
This overestimation comes from the shape of the tophat itself, which is a much broader distribution than either cubic spline or gaussian (at $|x|<h$).
Thus, the tophat kernel biases us towards preferentially larger contributions from particles.
}


\edit{
We therefore conclude that our key results of this work are largely insensitive to the choice of kernel weighting function.
}

\section{Dependence on Definition of ``Resolved'' Spaxel}
\label{appendix:cuts}

\begin{figure}
    \centering
    \includegraphics[width=\linewidth]{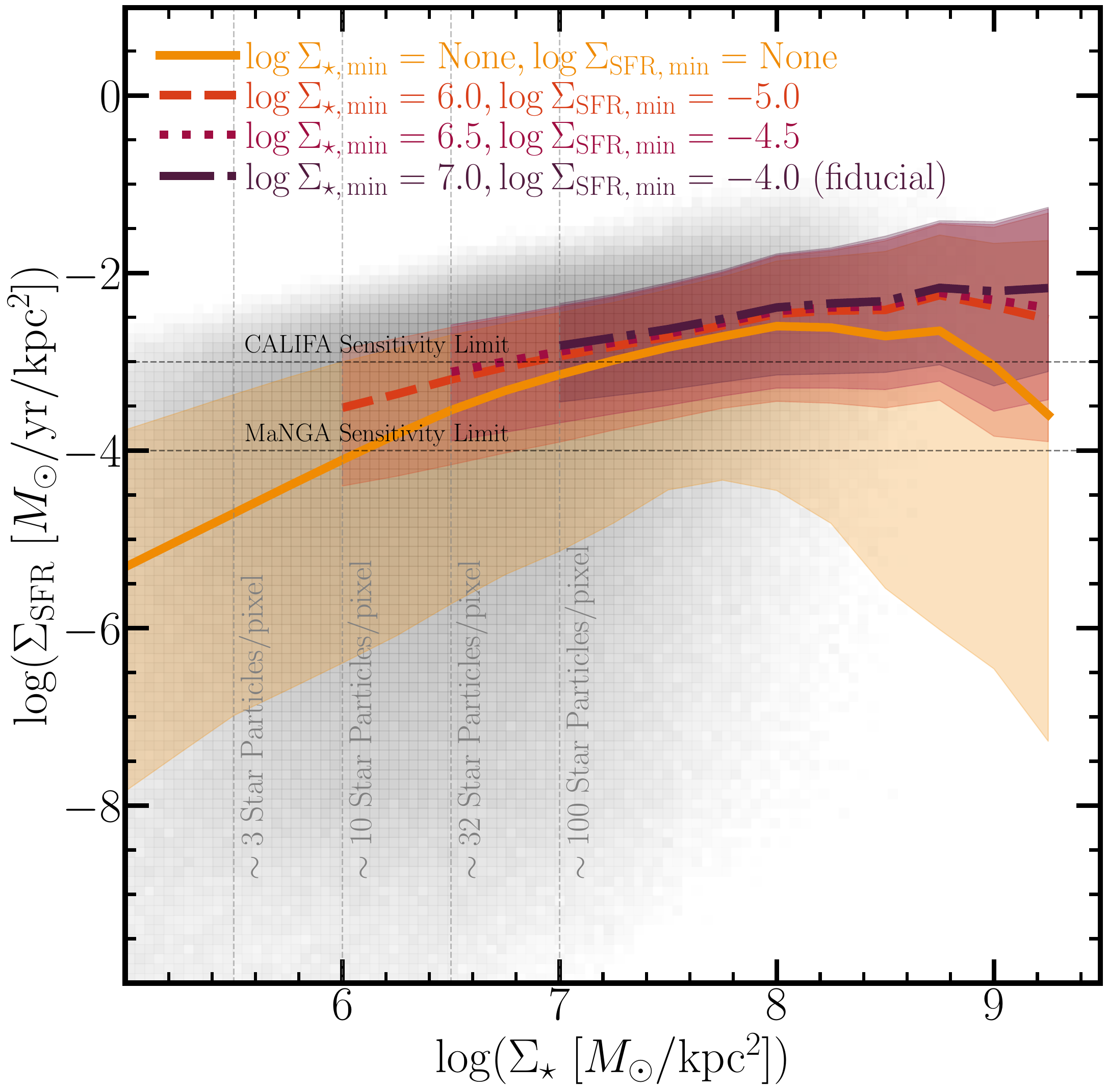}
    \caption{{\bf Dependence on Definition of ``Resolved'' Pixel}}
    \label{fig:resolved_pixel}
\end{figure}

\edit{
In Section~\ref{sec:method_map}, we outlined our pixel selection criteria which include a $\Sigma_\star$ cut for simulation resolution purposes and a $\Sigma_{\rm SFR}$ cut to ensure that our pixel would contain star formation (which is necessary to derive, e.g., emission line diagnostics for the metallicity).
Our fiducial cuts are $\Sigma_\star = 10^7~{M_\odot/{\rm kpc^2}}$, corresponding to roughly $100$ star particles per pixel, and $\Sigma_{\rm SFR}$, corresponding to a {\it specific} star formation rate surface density of $>10^{-11}~{\rm yr}^{-1}{\rm kpc}^{-2}$ and roughly in line with current observational limitations \citep{Diaz2016,Hsieh2017}.
In this appendix, we investigate the role these selection criteria play on our derived rSFMS.
}

\edit{
Figure~\ref{fig:resolved_pixel} shows the rSFMS using four different selection criteria:
(i) no minimum $\Sigma_{\star}$ or $\Sigma_{\rm SFR}$ solid orange line,
(ii) $\log(\Sigma_\star) > 6.0~M_\odot/{\rm kpc}^2$ and $\log(\Sigma_{\rm SFR}) > -5.0~M_\odot/{\rm yr}/{\rm kpc}^2$ dashed red line,
(iii) $\log(\Sigma_\star) > 6.5~M_\odot/{\rm kpc}^2$ and $\log(\Sigma_{\rm SFR}) > -4.5~M_\odot/{\rm yr}/{\rm kpc}^2$ dotted purple line,
(iv) $\log(\Sigma_\star) > 7.0~M_\odot/{\rm kpc}^2$ and $\log(\Sigma_{\rm SFR}) > -4.0~M_\odot/{\rm yr}/{\rm kpc}^2$ dot-dashed black line.
As a point of reference we include the full distribution of pixels in the background 2D histogram.
Moreover, we include vertical lines representing the contributions from approximately $3$, $10$, $32$, and $100$ star particles per pixel as well as horization lines showing the CALIFA \citep{Diaz2016} and MaNGA \citep{Hsieh2017} sensitivity limits.
}

\edit{
We find that being less restrictive in our selection criteria has only a moderate impact on our derived median rSFMS.
In particular, the median relations for selections (ii), (iii), and (iv) all agree within $\sim0.1$ dex.
The scatter of the distributions increases with decreasing minimum $\Sigma_{\rm SFR}$ due to the population of low $\Sigma_{\rm SFR}$ pixels that would otherwise be removed.
However, it should be noted that the power-law indices for the selection (ii), (iii), and (iv) are all in agreement within $\sim15\%$ ($0.35$, $0.31$, and $0.30$, respectively).
Selection criteria (i), on the other hand sees significant variation in terms of increased scatter (factor of $\sim2$ larger) and power law index ($0.54$, factor of $\sim1.5$ steeper).
All of the variation between the four different selection criteria is due to variation in the population of low $\Sigma_{\rm SFR}$ pixels.
}

\bibliography{main}{}
\bibliographystyle{aasjournal}

\end{document}